\documentclass[journal]{IEEEtran}
\usepackage[rgb,table,xcdraw]{xcolor}
\usepackage{hyperref}
\hypersetup{
    colorlinks=true,
    citecolor=blue,
}
\usepackage{colortbl}

\usepackage{graphicx}
\usepackage{cite}
\usepackage{amsmath}
\usepackage{amssymb}
\usepackage{pifont}
\usepackage{booktabs}
\usepackage{multirow}
\usepackage{diagbox}
\usepackage{subcaption}  

\newif\ifrevision
\revisionfalse

\definecolor{mygreen}{RGB}{0,100,0}

\ifrevision
  \newcommand{\rev}[1]{\textcolor{mygreen}{#1}}
\else
  \newcommand{\rev}[1]{#1}
\fi

\begin{document}
\title{MOS-Bench: Benchmarking Generalization Abilities of Subjective Speech Quality Assessment Models}

\author{Wen-Chin~Huang,~\IEEEmembership{Member,~IEEE,}
        Erica Cooper,~\IEEEmembership{Member,~IEEE,}
        and~Tomoki~Toda,~\IEEEmembership{Member,~IEEE}
\thanks{Wen-Chin Huang is with the Graduate School of Informatics, Nagoya University, Japan (e-mail: wen.chinhuang@g.sp.m.is.nagoya-u.ac.jp). Erica Cooper is with the National Institute of Information and Communications Technology (NICT), Japan. Tomoki Toda is with the Information Technology Center, Nagoya University, Japan.}
}

\markboth{Journal of \LaTeX\ Class Files,~Vol.~14, No.~8, August~2015}%
{Shell \MakeLowercase{\textit{et al.}}: Bare Demo of IEEEtran.cls for IEEE Journals}

\maketitle

\begin{abstract}
In this paper, we study the task of subjective speech quality assessment (SSQA), which refers to predicting the perceptual quality of speech. Owing to the development of deep neural network models, SSQA has greatly advanced and has been widely applied in scientific papers to evaluate speech generation systems. Nonetheless, the insufficient out-of-domain (OOD) generalization ability of current SSQA models is underexplored and often overlooked by researchers. To study this problem systematically, we present MOS-Bench, a diverse SSQA dataset collection that currently contains 8 training sets and 17 test sets. Through extensive experiments, we first highlight the OOD generalization challenges of existing models.
\rev{
We then evaluate the efficacy of multiple-dataset training, comparing straightforward data pooling against AlignNet, an existing domain-aware method. We demonstrate that pooling multiple training sets provides a simple yet effective solution, and variation in the data is a key factor for robust generalization beyond training data size.
}
\end{abstract}


\IEEEpeerreviewmaketitle

\section{Introduction}
\label{sec:intro}

\IEEEPARstart{S}{peech} quality assessment (SQA) refers to evaluating the quality of a speech sample \cite{speech-evaluation-review-2011, speech-evaluation-review-2024}. Since most ``end listeners'' of speech are humans, the most accurate way to assess quality is through subjective listening tests \cite{p800}. A common listening test protocol is the mean opinion score (MOS) test, in which the quality of a query speech sample is presented as the average of ratings from multiple listeners, usually on a five-point scale. However, conducting listening tests is inarguably very costly and time-consuming. Therefore, researchers have dedicated efforts to developing automated \textit{objective} evaluation methods, which can ease the evaluation process and accelerate the development iteration. In this work, we use the term \textbf{subjective speech quality assessment} (SSQA) to refer to the task of predicting the perceptual quality of speech.

The progress in the development of deep neural network (DNN) models has in turn greatly accelerated the development of \textit{model-based} objective methods for SSQA, which directly learn from datasets consisting of $\langle$speech $\textendash$ human rating$\rangle$ pairs. Researchers have explored various approaches, including different model architectures \cite{mosnet, wawenets, nisqa} and training objectives \cite{9414182, SQA-NMR-MOS, scoreq}. These efforts have advanced SSQA models to the point where they exhibit very strong correlations with human perception. For example, in the VoiceMOS Challenge (VMC) 2022 \cite{voicemos2022}, a scientific competition dedicated to SSQA, the best-performing system, UTMOS \cite{utmos}, achieved an exceptionally high correlation (0.959) with human ratings. Such results have encouraged speech generation researchers to increasingly adopt SSQA models in their work \cite{naturalspeech3, xtts, llama-omni}, and even to use them as primary evaluation metrics in scientific challenges \cite{dnsmos, dnsmos-p835, is2024-discrete, urgent2025, chime7-udase, chime7-evaluation}.

However, an underexplored aspect of current state-of-the-art SSQA models is their ability to generalize to out-of-domain (OOD) data, as first noted in \cite{ssl-mos}. In machine learning, a model exhibits good in-domain generalization when it performs well on test data drawn from the same distribution as the training data. In contrast, OOD generalization refers to scenarios in which test and training data come from different distributions. We argue that in SSQA, \textbf{almost all practical applications correspond to OOD settings}. Most SSQA models are trained on a single dataset derived from one listening test, which represents a unique context: distinct contents (text, speakers, etc.), recruited listeners, ranges of systems evaluated, and even instructions. During testing, incoming samples often come from, for example, a newly developed text-to-speech system or speech corrupted by previously unseen distortions. Such scenarios are naturally OOD.

A concrete demonstration of the limited OOD generalization ability of modern SSQA models comes from the results of VMC'23 \cite{voicemos2023}. The high correlation with human ratings observed in VMC'22 largely reflected an in-domain evaluation setting. In contrast, VMC'23 employed a fully zero-shot setup, where participants were asked to predict scores on three OOD test sets without access to any data from the corresponding domains. Consequently, most teams struggled to achieve strong performance across all sets. These results underscore the limitations of relying on current SSQA models as quality metrics in practical settings, such as those in scientific studies.

From these observations, we aim to systematically study the generalization ability of SSQA models in a standardized, large-scale setting. To this end, we introduce MOS-Bench, an open-sourced and ongoing collection of datasets for training and benchmarking subjective speech quality predictors. The currently released version includes eight training sets and 17 test sets, encompassing diverse sampling frequencies, languages, and speech types, ranging from the synthetic speech generated by text-to-speech (TTS), voice conversion (VC), and singing voice synthesis (SVS) systems to the naturally distorted or processed speech affected by noise, reverberation, codec artifacts, telephony, and more. Using each of the 8 training sets individually, we first benchmark the OOD generalization performance of state-of-the-art SSQA models across all 17 test sets, highlighting the challenges of OOD generalization.
\rev{
Building on this, we demonstrate the effectiveness of a simple yet overlooked approach: training on multiple datasets simultaneously. We compared a simple data-pooling approach against more complex, domain-aware architectures such as AlignNet \cite{alignnet}. Experimental results show that multiple dataset training substantially improves performance across all test sets, and that revealing data diversity, beyond dataset size, is crucial for robust generalization. Moreover, we show that the simple strategy of pooling diverse datasets yields superior and more robust performance. 
}
These findings provide practical guidance for constructing training corpora and developing future SSQA models with stronger OOD resilience.

\rev{
The contributions of this work are summarized as follows.
\begin{itemize}
\item \textbf{We introduce MOS-Bench, a standardized and reproducible benchmarking framework for SSQA}, to address the long-standing challenge of fragmented evaluation and enable consistent protocols across the research community.
\item \textbf{We conduct a large-scale systematic study of OOD generalization in SSQA}, providing a comprehensive analysis of the current state of the field, highlighting critical gaps in model robustness across diverse conditions and domains.
\item \textbf{We demonstrate the empirical effectiveness of multiple-dataset training}, revealing that dataset size alone does not determine OOD resilience, and provide indirect evidence that data diversity may play a more critical role. Our results further show that a straightforward data-pooling strategy can outperform specialized, complex architectures such as AlignNet, providing a high-performing and accessible baseline for future research.
\end{itemize}
}

\section{Related works}

\subsection{Benchmarking in subjective speech quality assessment}

In recent years, the field of SSQA has been driven forward by a surge in scientific challenges, with one of the most prominent being the VoiceMOS Challenge series \cite{voicemos2022, voicemos2023, voicemos2024}, recently renamed as the AudioMOS Challenge \cite{audiomos2025}. Before these challenges emerged, SSQA datasets were mostly proprietary, making it difficult to compare results across studies because of differences in data collection and evaluation procedures. In contrast, these challenges provided standardized datasets and evaluation protocols, which enabled direct comparisons and reproducibility. The VoiceMOS Challenge series, in particular, stimulated innovation by attracting more than 40 participating teams over three consecutive years, and its datasets have since been widely reused in follow-up research. Similar initiatives have also been launched, such as the ConferencingSpeech Challenge 2022 \cite{conferencingspeech2022}, which focused on online conferencing scenarios, and the URGENT Challenge 2026 \cite{urgent2026}, which emphasized the assessment of enhanced speech.

Despite these advances, scientific challenges inherently constrain progress by fixing the dataset and evaluation protocol, thereby narrowing the scope of model development. This setting encourages system optimization toward the challenge data but often leads to overfitting to particular corpora. As a result, models that perform strongly in these tracks may struggle to generalize to unseen datasets with different domains, languages, or distortion types, as we saw in VMC'23. These observations highlight the need for benchmarks that explicitly emphasize OOD generalization, such as MOS-Bench, as well as models and evaluation methodologies tailored to assessing and improving cross-dataset robustness.

\subsection{Improving generalization ability in speech quality assessment}

In this subsection, we discuss techniques that have been applied to improving the generalization ability of SSQA. The first technique is \textbf{transfer learning}, where knowledge learned from a pretext task is utilized on the task at hand. A representative method is self-supervised learning (SSL), which has been shown to be extremely effective in a wide range of speech processing tasks \cite{speech-ssl-review}. First, pretraining is conducted on a large-scale unlabeled speech dataset w.r.t. some pretext task, such as contrastive learning \cite{wav2vec2} or masked prediction \cite{hubert, wavlm}, followed by fine-tuning with a labeled dataset for the downstream task. Although there have been many attempts to apply SSL to the task of SSQA \cite{first-ssl-mos, squid, mosanet, sqa-xlsr}, perhaps the most representative approach is SSL-MOS \cite{ssl-mos}, where the authors applied a very simple model architecture by attaching a linear head onto SSL models. SSL-MOS was shown to not only improve performance on in-domain datasets but also provide reasonable correlations in the zero-shot setting.

The second technique is \textbf{ensemble learning}, which aims to reduce model variance and improve robustness by combining multiple models. For instance, UTMOS \cite{utmos}, which was the top-performing team in VMC'22, performed stacking \cite{stacking} with a set of strong and weak learners, where they employed regression models, including linear regression, decision trees, and kernel methods as the weak learners. LE-SSL-MOS \cite{le-ssl-mos}, the top performing team in VMC'23, combined scores from multiple models, including a vanilla SSL-MOS model, an SSL-MOS model with listener modeling, a SpeechLMScore model \cite{speechlmscore}, and the confidence score of an ASR model. LE-SSL-MOS was the only team in VMC'23 that performed well on all tracks, demonstrating its strong generalization ability.

Perhaps the most straightforward way to increase the generalization ability is \textbf{multiple-dataset training} by simply pooling several listening test datasets into one large training set. In fact, this multiple-dataset training has been adopted in some research papers \cite{nisqa, utmosv2, sq-ast}. However, this approach has an issue regarding the phenomenon called the ``\textit{corpus effect}'' as defined in \cite{p1401}. The corpus effect refers to the problem of the same type of speech receiving different scores on different listening tests. For instance, it was reported that the same TTS system received a lower score when systems with higher qualities were present in the listening test \cite{back-to-the-future}. This is because MOS can be affected by listener preferences and the range of conditions included in one listening test. The latter is sometimes called the ``range-equalizing bias'' \cite{range-equalizing-bias}. A pioneering attempt to solve this problem is the bias-aware loss \cite{biaw-aware-loss}, which attempts to assign different weights to different datasets, with the cost of a complex training process. Recently, AlignNet has been proposed \cite{alignnet}, with a much simpler training process than the bias-aware loss but higher performance.


\begin{table*}[t]

\footnotesize

\centering
\caption{
\rev{
Details of the training and development sets in MOS-Bench. 
Abbreviations: En. = English, Ch. = Chinese, Ja. = Japanese, fs = sampling frequency, SVS = singing voice synthesis, SVC = singing voice conversion, VC = voice conversion, TTS = text-to-speech, PSTN = public switched telephone network. 
An asterisk (*) in the interlistener variance column indicates that samples with only one rating were assigned variance = 0, so the actual variance may be higher than reported.
The listener-wise scores were not provided for Tencent, so the interlistener variance could not be calculated.
}
}

\label{tab:mos-bench-train}

\begin{tabular}{@{}c|c|c|c|c|c|cc|c|c@{}}
\toprule
\multirow{2}{*}{Type} & \multirow{2}{*}{Name} & \multirow{2}{*}{Domain} & \multirow{2}{*}{Lang.} & \multirow{2}{*}{\begin{tabular}[c]{@{}c@{}}fs\\(kHz)\end{tabular}} & \multirow{2}{*}{\begin{tabular}[c]{@{}c@{}}\# Samples\\ (train/dev)\end{tabular}} & \multicolumn{2}{c|}{\# rating/sample} & \multirow{2}{*}{\begin{tabular}[c]{@{}c@{}}Inter-listener\\ variance\\ (mean ± std)\end{tabular}} & \multirow{2}{*}{\begin{tabular}[c]{@{}c@{}}Provided\\listener\\metadata\end{tabular}} \\
\cmidrule(lr){7-8}
 &  &  &  &  &  & min/max & mean ± std &  \\[5pt]
\midrule
\multirow{3}{*}{Synthetic} & BVCC & TTS, VC, clean speech & En. & 16 & 4944/1066 & 8/8 & 8.0 ± 0.0 & 0.63 ± 0.36 & \begin{tabular}[c]{@{}c@{}}ID, gender,\\age range,\\hearing impairment\end{tabular} \\ \cmidrule(l){2-10} 
 & SOMOS & TTS, clean speech & En. & 24 & 14100/3000 & 4/19 & 10.7 ± 2.1 & 1.28 ± 0.53 & \begin{tabular}[c]{@{}c@{}}ID,\\English locale\end{tabular} \\ \cmidrule(l){2-10} 
 & SingMOS & \begin{tabular}[c]{@{}c@{}}SVS, SVC,\\clean singing voice\end{tabular} & Ch., Ja. & 16 & 2000/544 & 5/5 & 5.0 ± 0.0 & 0.14 ± 0.22 & ID \\ \midrule
\multirow{5}{*}{Distorted} & NISQA & \begin{tabular}[c]{@{}c@{}}simulated distorted speech,\\real distorted speech,\\clean speech\end{tabular} & En. & 48 & 11020/2700 & 3/10 & 5.2 ± 1.1 & 0.54 ± 0.43 & -- \\ \cmidrule(l){2-10} 
 & TMHINT-QI & \begin{tabular}[c]{@{}c@{}}simulated noisy speech,\\enhanced speech,\\clean speech\end{tabular} & Ch. & 16 & 11644/1293 & 1/6 & 1.3 ± 0.5 & 0.16 ± 0.45* & ID \\ \cmidrule(l){2-10} 
 & Tencent & \begin{tabular}[c]{@{}c@{}}simulated distorted speech,\\clean speech\end{tabular} & Ch. & 16 & 10408/1155 & \multicolumn{2}{c|}{\textgreater{}20} & -- & -- \\ \cmidrule(l){2-10} 
 & PSTN & \begin{tabular}[c]{@{}c@{}}PSTN calls,\\simulated distorted speech\end{tabular} & En. & 8 & 52839/5870 & 1/10 & 4.6 ± 1.6 & 0.80 ± 0.72* & -- \\ \cmidrule(l){2-10} 
 & \begin{tabular}[c]{@{}c@{}}URGENT\\ 2024-MOS\end{tabular} & \begin{tabular}[c]{@{}c@{}}artificial \& real\\distorted speech,\\ enhanced speech\end{tabular} & En. & 8-48 & 6210/690 & 8/8 & 8.0 ± 0.0 & 0.87 ± 0.48 & -- \\ \bottomrule
\end{tabular}

\end{table*}


\begin{table*}[t]


\centering
\caption{Details of the test sets in MOS-Bench. 
Abbreviations: En. = English, Ch. = Chinese, Ja. = Japanese, Fr. = French, Pt. = Portuguese, Nl. = Dutch, fs = sampling frequency, SVS = singing voice synthesis, SVC = singing voice conversion, VC = voice conversion, TTS = text-to-speech, VoIP = voice over Internet protocol. 
The listener-wise scores were not provided for TCD-VOIP, so statistics could not be calculated.
An asterisk (*) in the interlistener variance column indicates that samples with only one rating were assigned variance = 0, so the actual variance may be higher than reported.}
\label{tab:mos-bench-test}

\begin{tabular}{@{}c|c|c|c|c|c|cc|c@{}}
\toprule
\multirow{2}{*}{Type} & \multirow{2}{*}{Name} & \multirow{2}{*}{Domain} & \multirow{2}{*}{Lang.} & \multirow{2}{*}{\begin{tabular}[c]{@{}c@{}}fs\\ (kHz)\end{tabular}} & \multirow{2}{*}{\# Samples} & \multicolumn{2}{c|}{\# rating/sample} & \multirow{2}{*}{\begin{tabular}[c]{@{}c@{}}inter-listener\\ variance\\ (mean ± std)\end{tabular}} \\ \cmidrule(lr){7-8}
 &  &  &  &  &  & min/max & mean ± std &  \\[5pt]
 \midrule
\multirow{10}{*}{Synthetic} & BVCC & TTS, VC, natural speech & En. & 16 & 1066 & 8/8 & 8.0 ± 0.0 & 0.61 ± 0.35 \\ \cmidrule(l){2-9} 
 & SOMOS & TTS, natural speech & En. & 24 & 3000 & 4/19 & 10.8 ± 2.2 & 1.28 ± 0.53 \\ \cmidrule(l){2-9} 
 & BC19 & TTS, natural speech & Ch. & 16 & 540 & 10/17 & 14.2 ± 1.7 & 0.78 ± 0.39 \\ \cmidrule(l){2-9} 
 & BC23 a & TTS, natural speech & Fr. & 22 & 882 & 15/20 & 17.3 ± 1.5 & 0.80 ± 0.35 \\ \cmidrule(l){2-9} 
 & BC23 b & TTS, natural speech & Fr. & 22 & 578 & 15/20 & 16.8 ± 1.3 & 0.82 ± 0.39 \\ \cmidrule(l){2-9} 
 & SVCC23 & SVC, natural singing voice & En. & 24 & 4040 & 5/26 & 6.3 ± 2.1 & 0.77 ± 0.52 \\ \cmidrule(l){2-9} 
 & SingMOS & SVS, SVC, natural singing voice & Ch., Ja. & 16 & 645 & 5/10 & 6.1 ± 2.1 & 0.18 ± 0.25 \\ \cmidrule(l){2-9} 
 & BRSpeechMOS & TTS, natural speech & Pt & 16 & 243 & 1/20 & 2.0 ± 2.7 & 0.23 ± 0.53* \\ \cmidrule(l){2-9} 
 & HablaMOS & TTS, natural speech & Es. & 16 & 408 & 1/2 & 1.1 ± 0.1 & 0.01 ± 0.05* \\ \cmidrule(l){2-9} 
 & TTSDS2 & TTS & En. & 22 & 4731 & 7/34 & 10.3 ± 4.0 & 0.97 ± 0.43 \\ \midrule
\multirow{7}{*}{Distorted} & NISQA FOR & artificial distorted speech, VoIP & En. & 48 & 240 & 21/34 & 20.4 ± 3.6 & 0.50 ± 0.20 \\ \cmidrule(l){2-9} 
 & NISQA LIVETALK & real-world distorted speech, VoIP & Nl. & 48 & 232 & 24/34 & 24.0 ± 0.0 & 0.66 ± 0.28 \\ \cmidrule(l){2-9} 
 & NISQA P501 & artificial distorted speech, VoIP & En. & 48 & 240 & 18/34 & 28.3 ± 3.6 & 0.42 ± 0.23 \\ \cmidrule(l){2-9} 
 & TMHINT-QI & \begin{tabular}[c]{@{}c@{}}artificial noisy speech,\\ enhanced speech, clean speech\end{tabular} & Ch. & 16 & 1978 & 3/11 & 3.5 ± 1.1 & 0.64 ± 0.62 \\ \cmidrule(l){2-9} 
 & TMHINT-QI (S) & \begin{tabular}[c]{@{}c@{}}artificial noisy speech,\\ enhanced speech, clean speech\end{tabular} & Ch. & 16 & 1960 & 4/7 & 5.3 ± 1.0 & 0.74 ± 0.55 \\ \cmidrule(l){2-9} 
 & TCD-VOIP & artificial distorted speech, VoIP & En. & 48 & 384 & \multicolumn{2}{c|}{24} & -- \\ \cmidrule(l){2-9} 
 & VMC24 track3 & \begin{tabular}[c]{@{}c@{}}artificial noisy speech,\\ enhanced speech, clean speech\end{tabular} & En. & 16 & 280 & 5/5 & 5.0 ± 0.0 & 0.38 ± 0.29 \\ \bottomrule
\end{tabular}

\end{table*}


\section{Description of MOS-Bench}

\subsection{General description}
\label{ssec:general-description-mos-bench}

The MOS-Bench collection contains two parts: the 8 training and development sets, as summarized in Table~\ref{tab:mos-bench-train}, and the 17 test sets, as summarized in Table~\ref{tab:mos-bench-test}.
All datasets included in MOS-Bench are publicly available and can be easily accessed using the SHEET toolkit \cite{sheet}.
The listening test protocol in each dataset was a MOS test with scores ranging from 1 to 5.
Although most datasets come with a train/dev/test split, for those without an official split, we randomly divided them into a train/dev set with a 9:1 ratio.

The MOS-Bench demonstrates a diverse collection of datasets with different properties. In this paper, we roughly divide speech into two types:
\begin{itemize}
    \item \textbf{Synthetic speech}: speech generated by, but not limited to, TTS, VC, SVS, and SVC systems
    \item \textbf{Distorted speech}: speech that underwent various distortions, including artificially added and real noise, reverberation, voice-over-IP (VoIP), codec, and replay
\end{itemize}
The main difference is that for the distorted speech, there exists a ground truth to be compared with, typically the speech before undergoing the distortion. The synthetic speech, on the other hand, often does not come with a clear ground truth owing to the one-to-many problem in speech synthesis. In addition, MOS-Bench covers a broad spectrum of languages: English, Chinese, Japanese, French, Spanish, Dutch, and Brazilian-Portuguese. MOS-Bench also covers a wide range of sampling frequencies, ranging from 8 kHz to 48 kHz.

\subsection{Brief descriptions of each dataset}

In this subsection, we briefly describe the properties of each dataset. For details, please refer to the respective original papers.

\subsubsection{BVCC}

The BVCC dataset \cite{bvcc} is a large-scale listening test covering English samples from 187 different TTS and VC systems, which mainly come from past years of the Blizzard Challenges (BC) and Voice Conversion Challenges (VCC), as well as published samples from ESPnet-TTS \cite{espnet-tts}.
We used the official partition, where the dev and test sets contain unseen synthesis systems, speakers, texts, and listeners not in the training set, whereas the rating distribution of each set is matched as closely as possible.

\subsubsection{SOMOS}

The SOMOS dataset \cite{somos} contains samples from 200 neural TTS systems, which were all trained on LJSpeech \cite{ljspeech}, a single-speaker English female corpus, with a shared LPCNet vocoder \cite{lpcnet}. These TTS systems differ mainly in the acoustic model architecture, as well as input variations, including syntactic, linguistic, and prosodic clues. In addition, clean speech samples were also included.
We used the official train/dev/test partition within the \texttt{clean} set.

\subsubsection{SingMOS}

The SingMOS dataset \cite{singmos} focuses on singing voices and includes samples collected from 35 modern neural SVS and SVC systems, including natural singing voice samples. These systems are trained on open-source Chinese and Japanese singing voice datasets.
The SingMOS dataset was also used in VMC'24 track 2, and we used the official train/dev/test split in the challenge.

\subsubsection{NISQA}

The NISQA dataset \cite{nisqa} was designed to evaluate speech with distortions occurring in communication networks. In MOS-Bench, the train and dev sets combined the \texttt{NISQA TRAIN SIM}, \texttt{NISQA TRAIN LIVE} sets, and the \texttt{NISQA VAL SIM}, \texttt{NISQA VAL LIVE} sets, respectively.
The \texttt{NISQA TRAIN SIM} and \texttt{NISQA VAL SIM} sets contained (1) simulated speech distortions, such as packet loss, bandpass filter, different codecs, and clipping, and (2) artificially added background noises.
The \texttt{NISQA TRAIN LIVE} and \texttt{NISQA VAL LIVE} sets contained live Skype and phone recordings, with real distortions created during recording, such as keyboard typing and street noise. 

The \texttt{NISQA TEST P501}, \texttt{NISQA TEST FOR}, and \texttt{NISQA TEST LIVETALK} sets were used as the test sets in MOS-Bench\footnote{The \texttt{NISQA TEST NSC} set was excluded since it was not publicly available anymore.}. The \texttt{NISQA TEST P501} and \texttt{NISQA TEST FOR} sets included samples with simulated distortions, as well as live VoIP calls where speech samples were played back directly from a laptop, followed by a simulated packet loss, warping, a low bitrate, etc.
Finally, \texttt{NISQA TEST LIVETALK} is a German dataset with real phone call recordings, where talkers spoke directly into the terminal device in different backgrounds with different distortions. Talkers can be called either through the mobile network or via VoIP.

\subsubsection{TMHINT-QI}

The TMHINT-QI dataset \cite{tmhintqi} contained Taiwanese Mandarin samples with four added artificial noise types (babble, street, pink, and white) at four signal-to-noise (SNR) ratio levels (-2, 0, 2, and 5), along with their enhanced versions using five speech enhancement systems. We used the official train/dev/test split. Note that all noise types, SNR levels, and enhancement systems are the same across the splits.

\subsubsection{Tencent}

The Tencent dataset \cite{conferencingspeech2022} was a Chinese dataset designed for speech quality assessment in online conference scenarios. First, samples from three publicly available datasets were artificially distorted with either background noise, clipping, or codec compression loss. Then, noise suppression and packet loss concealment were applied to simulate a realistic online conference scenario. In addition, samples with either simulated reverberation or recorded in a reverberant environment were also considered, each with a different room size and reverberation delay. Most samples have a sampling frequency of 16 kHz, whereas some are at 48 kHz.
As part of the dataset of the ConferencingSpeech2022 challenge \cite{conferencingspeech2022}, the organizers only provided the training partition. Therefore, we used a random 9:1 train/dev split.

\subsubsection{PSTN}

The PSTN dataset \cite{pstn} was collected to study the speech quality of a public switched telephone network. Similar to the Tencent dataset, the PSTN dataset was also part of the ConferencingSpeech2022 challenge, so we used the filtered version provided by the organizers and used a random 9:1 train/dev split. The PSTN dataset contains automatically conducted phone calls between a PSTN and a VoIP endpoint. The 500,000 original automated calls were further filtered to maintain a reasonable quality distribution. Then, background noises were artificially added to simulate real-world phone calls.

\subsubsection{URGENT2024-MOS}

The URGENT2024-MOS dataset \cite{urgent2024, urgent-p808} was part of the dataset of the URGENT challenge 2024, a challenge designed to evaluate the performance of speech enhancement systems in handling various distortions and input conditions. The URGENT2024-MOS dataset consists of the original noisy and 22 enhanced versions of a subset (300 samples) of the blind test set, resulting in 6900 speech samples in total. The original noisy subset contained both artificial and real distorted English monaural samples with sampling frequencies ranging from 8 kHz to 48 kHz. The ratings were collected following the P.808 \cite{p808} recommendation using the Amazon Mechanical Turk platform. The raw ratings from each subject were averaged to obtain the final MOS score.

\subsubsection{BC19}

The Blizzard Challenge 2019 \cite{blizzard2019} listening test data served as the OOD track of VMC 2022. This dataset contains Chinese TTS samples.

\subsubsection{BC23 a and BC23 b}

The BC 2023 \cite{blizzard2023} listening test data served as tracks 1a and 1b of VMC 2023. The task setting in the Blizzard Challenge 2023 was French TTS, with two subtasks: the main Hub task provided 50 h of training data from a female speaker, and the Spoke task was a speaker adaptation task where 2 h of data from a different female speaker was provided. Two different listening tests were conducted separately on the submitted TTS samples for these two tasks.

\subsubsection{SVCC23}

The Singing Voice Conversion Challenge (SVCC) 2023 \cite{svcc2023} listening test data served as track 2 of VMC 2023. SVCC'23 focused on singing voice conversion, where there were two tasks: an in-domain task, where the singing voice samples of the target were provided, and a cross-domain task, where the normal voice samples were provided. Each sample was rated by six listeners.

\subsubsection{BRSpeechMOS}

The BRSpeechMOS dataset \cite{brspeechmos} contains Brazilian$\textendash$Portuguese TTS and natural samples at 16 kHz. Unfortunately, very little metadata information, such as the number of speakers and TTS systems, could be found. From manual inspection, the dataset seems to contain clean and noisy natural samples, in addition to samples from two neural TTS systems.

\subsubsection{HablaMOS}

The HablaMOS dataset \cite{hablamos} consists of 16 kHz single-speaker TTS samples from 52 different systems, as well as natural samples. These systems vary in gender, accent, and region. 100 text samples were randomly sampled from the Argentinian Spanish speech dataset \cite{argentinian-spanish-speech-dataset}. In addition, vocal tract length perturbation and phase alteration were also applied to increase the variation of the dataset. The listening test was based on the ITU-T Rec. P.807 standard \cite{p807}, and a total of 92 listeners participated.

\subsubsection{TTSDS2}

The TTSDS2 dataset \cite{ttsds2} contains samples from 20 TTS systems from 2022 to 2024 whose quality was claimed to reach human parity. The materials cover four domains: clean and noisy audiobooks, in-the-wild YouTube speech, and children’s dialogue. A crowdsourcing listening test was conducted with a total of 200 native listeners.

\subsubsection{TMHINT-QI (S)}

The TMHINT-QI(S) \cite{tmhintqi-s} dataset served as track 3 in VMC 2023. The base speech samples and the noisy utterance generation process were the same as those in the TMHINT-QI dataset. Two out of five enhancement methods in TMHINT-QI(S) were different from those in TMHINT-QI, resulting in five partially different enhancement methods. In addition, a separate listening test with non-overlapping listeners was conducted.

\subsubsection{TCD-VOIP}

The TCD-VOIP dataset \cite{tcd-voip} was created to study speech quality degradations typical in VoIP applications. It contains speech distorted by five platform-independent degradations: background noise, competing speakers, echo effects, amplitude clipping, and choppy speech (simulated packet loss). The base speech samples are from the TSP speech database \cite{tsp} with four speakers (two males and two females) reading Harvard sentences \cite{harvard-sentences}. 24 native, normal hearing listeners participated in the listening test.

\subsubsection{VMC24 track3}

The VMC24 track 3 dataset \cite{voicemos2024} was based on the VoiceBank-DEMAND \cite{voicebank-demand} dataset. First, 40 clean utterances were contaminated by one of the four types of noise at an SNR of 0 dB. Then, each noisy utterance was processed by five speech enhancement systems, resulting in a total of 280 utterances. A total of ten listeners (six males and four females, all native English speakers) participated in the listening test.

\subsection{Dataset quality assessment}
\label{ssec:dataset-quality-assessment}

In recent years, the MOS test has faced criticism \cite{back-to-the-future, kirkland23_ssw, limits-of-mos}, particularly regarding inconsistencies in its design across different studies. Researchers have emphasized the importance of carefully reporting listening test details \cite{report-details-in-subjective-evaluation, good-practices}, as factors such as the number of ratings and the type of listener can strongly affect the reliability of the results \cite{are-we-using-enough-listeners, mos-tail-prob-analysis}. Assessing the quality of the listening test datasets in MOS-Bench is challenging, as we are aggregating existing datasets and have no control over missing or incomplete metadata.

Therefore, in this subsection, we report statistics of two types of listening test metadata. First, we report the \textbf{minimum, maximum, mean, and standard deviation} of the \textbf{number of ratings per sample}, as shown in Tables~\ref{tab:mos-bench-train} and \ref{tab:mos-bench-test}. Although there is no clear guideline for how many ratings per sample are ``enough'', many datasets contain samples with extremely few ratings -- some even with just one. We also observed that several datasets have an average of two ratings or fewer per sample (TMHINT-QI training set, BRSpeechMOS, and HablaMOS), and some datasets exhibit a standard deviation larger than two (SOMOS train and test sets, SVCC23, SingMOS, BRSpeechMOS, TTSDS2, NISQA FOR, and NISQA P501), a natural consequence of crowdsourced listening tests where some samples receive far fewer ratings than others, creating an imbalance in reliability across samples.

Next, we report the \textbf{mean and standard deviation} of a quantity we call the \textbf{interlistener variance}. This quantity is designed to measure the agreement across listeners. Specifically, in an SSQA dataset where each sample may receive a different number of ratings, we first determine, for each sample, how spread out the ratings are by computing the variance across its listeners. We refer to this value as the \textit{interlistener variance}. Finally, we report the mean and standard deviation of these interlistener variances across the entire dataset.

From Tables~\ref{tab:mos-bench-train} and \ref{tab:mos-bench-test}, we find that the mean interlistener variance for most datasets is below one, with the exception of the SOMOS train and test sets. To interpret this value, note that the maximum possible interlistener variance is four, which occurs when half of the listeners rate a sample as one and the other half as five. A variance of one, by contrast, corresponds to a situation where ratings are closer together (for instance, half of the listeners rate two and the other half rate four). Thus, on average, listener agreement is high in most datasets. However, the standard deviation of the interlistener variance is also relatively large, indicating that whereas many samples yield consistent judgments, others trigger substantial disagreement among listeners.

Note that each of the above-mentioned two statistics alone cannot serve as the sole indicator of dataset quality. For instance, if a dataset has a reasonable number of listeners and still a high variance, instead of being considered a low-quality dataset, it may be considered a challenging dataset, as listeners may reasonably disagree for good reasons. We would also like to emphasize that in this subsection, we focus on the critical analysis of dataset quality. Given the limited number of available SSQA datasets, excluding datasets with lower quality would likely cause more harm than benefit.

Note that neither of the two statistics discussed above can serve as a standalone indicator of dataset quality. For example, a dataset with a sufficient number of listeners but a high variance should not be regarded as a low-quality dataset; rather, it may simply be more \textit{challenging}, reflecting reasonable disagreement among listeners. We also emphasize that in this subsection, we provide only a critical assessment of dataset quality. Given the limited availability of SSQA datasets, excluding those as being of ``lower quality'' would likely cause more harm than benefit.

\begin{figure}[t]
	\centering
        \includegraphics[width=\linewidth]{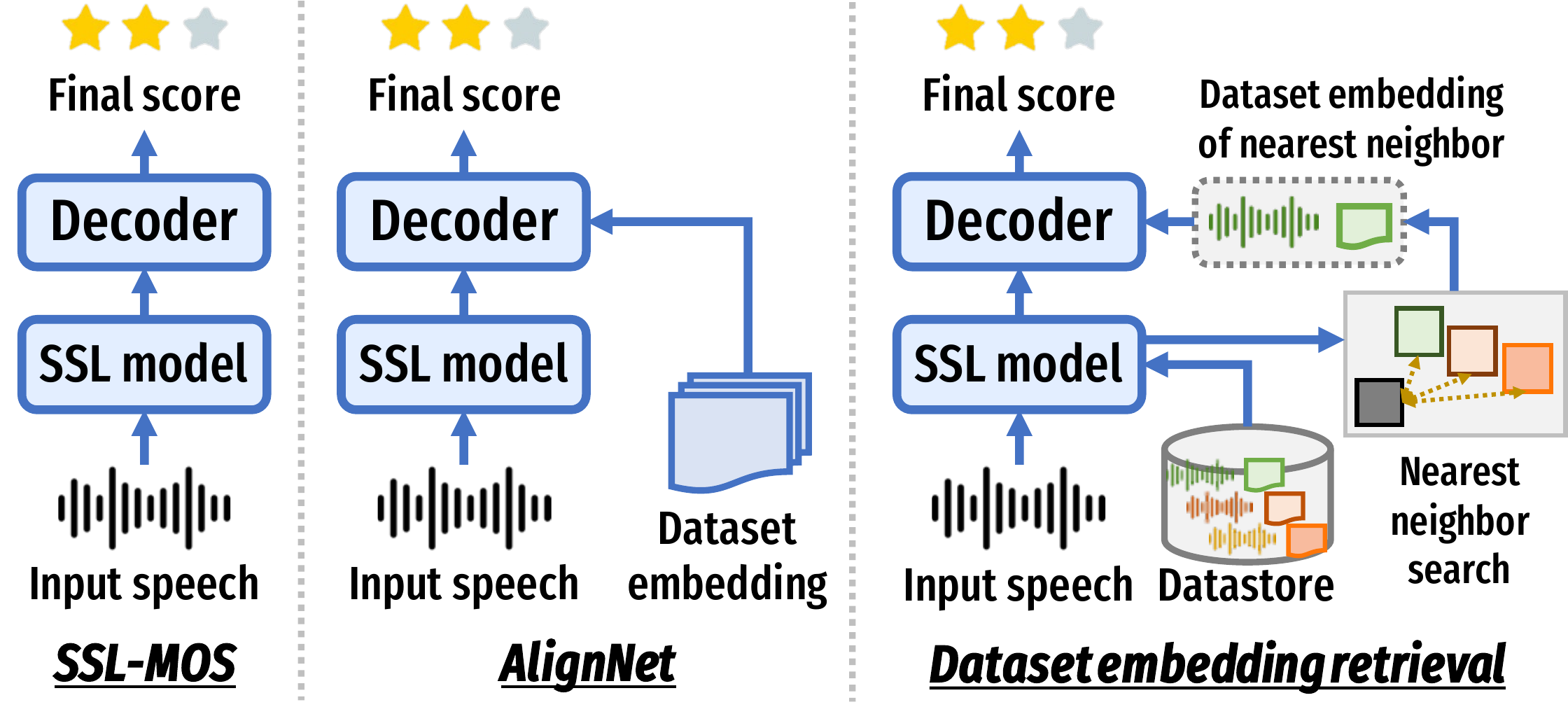}
	\caption{\label{fig:models}Models and inference methods used in this work. The time pooling operation is omitted.}
\end{figure}

\section{Models}


\subsection{Main model backbone: SSL-MOS}

The main model backbone adopted in most experiments in this work was SSL-MOS \cite{ssl-mos}, one of the most popular architectures in SSQA, because of its simplicity and excellent performance.
The left side of Fig.~\ref{fig:models} illustrates SSL-MOS. Given an input speech $x$, the SSL model $\textsc{SSL}(\cdot)$ outputs a sequence of frame-wise hidden representations, which is further sent into the decoder $\textsc{Dec}(\cdot)$ to generate a frame-wise score sequence. The predicted score $\hat{y}$ is then obtained by time pooling over the frame-wise score sequence:
\begin{equation}
    \hat{y} =\text{TimePooling}(\textsc{Dec}(\textsc{SSL}(x))).
\end{equation}
Finally, the training objective is an L1 loss between the ground truth score $y$ and the predicted score $\hat{y}$.
During training, $\textsc{SSL}(\cdot)$ and $\textsc{Dec}(\cdot)$ are jointly optimized. During inference, the process to obtain the predicted score $\hat{y}$ is the same as that during training.

\subsection{AlignNet}
\label{ssec:alignnet}

The AlignNet was a model designed to tackle the potential corpus effect in multi-dataset training \cite{alignnet}, as illustrated in the middle of Fig.~\ref{fig:models}. The core idea is similar to listener modeling \cite{mbnet, ldnet}, which is to have the encoder behave in a dataset-independent fashion, whose output is then mapped to the dataset-specific score by the decoder.

\subsubsection{Original formulation}

Suppose we combine $K$ training sets to form a large training set, $\mathcal{D}_{\text{train}}$, where each training sample is associated with the dataset ID $d(x)$, denoting which training set it originally belonged to: $d(x)\in\{1,\dots,K\}$. The AlignNet model consists of an encoder and a decoder, where we use the same network architecture as that of SSL-MOS. In addition, a set of learnable dataset embeddings is maintained: $\mathcal{E}=\{e_1,\dots,e_K\}$.

In the forward pass, the only difference between AlignNet and SSL-MOS is that the decoder $\textsc{Dec}(\cdot,\cdot)$ now takes two inputs: the encoder output and a \textit{dataset embedding}. The encoding and decoding processes are identical to those of SSL-MOS, and the predicted score is obtained as
\begin{equation}
    \hat{y} =\text{TimePooling}(\textsc{Dec}(\textsc{SSL}(x), e_{d(x)})).
\end{equation}

During training, for each training sample, the dataset ID is already known, so we can directly use it. However, during inference, we need to decide which dataset embedding to use. The solution proposed in the original AlignNet paper was to select a \textit{reference dataset} from all the training datasets beforehand. During training, for all the samples in the reference dataset, the decoder is forced to become the identity function. Then, during inference, the embedding of the reference dataset is used, regardless of the type of input speech. Such an approach can also ground the encoder to produce meaningful quality scores. However, the behavior of AlignNet will largely depend on the choice of the reference dataset. For instance, the experiments described in the original paper used NISQA as the reference dataset; therefore, there would be an obvious mismatch when the input speech is from a TTS system.

\subsubsection{Proposed inference method: dataset embedding retrieval}

We therefore propose a new inference method called \textit{dataset embedding retrieval}, illustrated on the right side of Fig.~\ref{fig:models}, which is inspired by recent retrieval-based methods in SQA \cite{ramp}.
After the model training phase, we construct a \textit{datastore}, $\mathcal{R}$, as follows:
\begin{align}
    h(x) &= \text{TimePooling}(\textsc{SSL(x)}),\\
    \mathcal{R} &= \bigl\{
        \bigr(
            h(x), d(x)
        \bigr) ~ | ~ x\in\mathcal{D}_{\text{train}}
    \bigr\},
\end{align}
where each pair contains $h(x)$, the time-pooled SSL representations of each training sample $x$, and $d(x)$, the dataset ID associated with it. 
During inference, given the input test sample $x_{\text{test}}$, we first extract its time-pooled SSL representation, $h(x_{\text{test}})$. Then, from the datastore, we retrieve the nearest neighbor to $h(x_{\text{test}})$ w.r.t. some predefined distance measure $\textsc{dist}(\cdot,\cdot)$:
\begin{equation}
    (h^*,d^*)=\arg\min_{(h_i,d_i)\in \mathcal{R}} \textsc{dist}(h_i, h(x_{\text{test}})).
\end{equation}
We may then use the dataset ID to decide the retrieved dataset embedding $e_{d^*}$.
Here, we use the negative cosine similarity as the distance measure. Finally, the predicted score $\hat{y}$ can be obtained by retrieved dataset embedding:
\begin{equation}
    \hat{y} =\text{TimePooling}(\textsc{Dec}(\textsc{SSL}(x), e_{d^*}).
\end{equation}

\begin{figure*}[t]
  \centering
  \begin{subfigure}{0.49\textwidth}
    \centering
    \includegraphics[width=\linewidth]{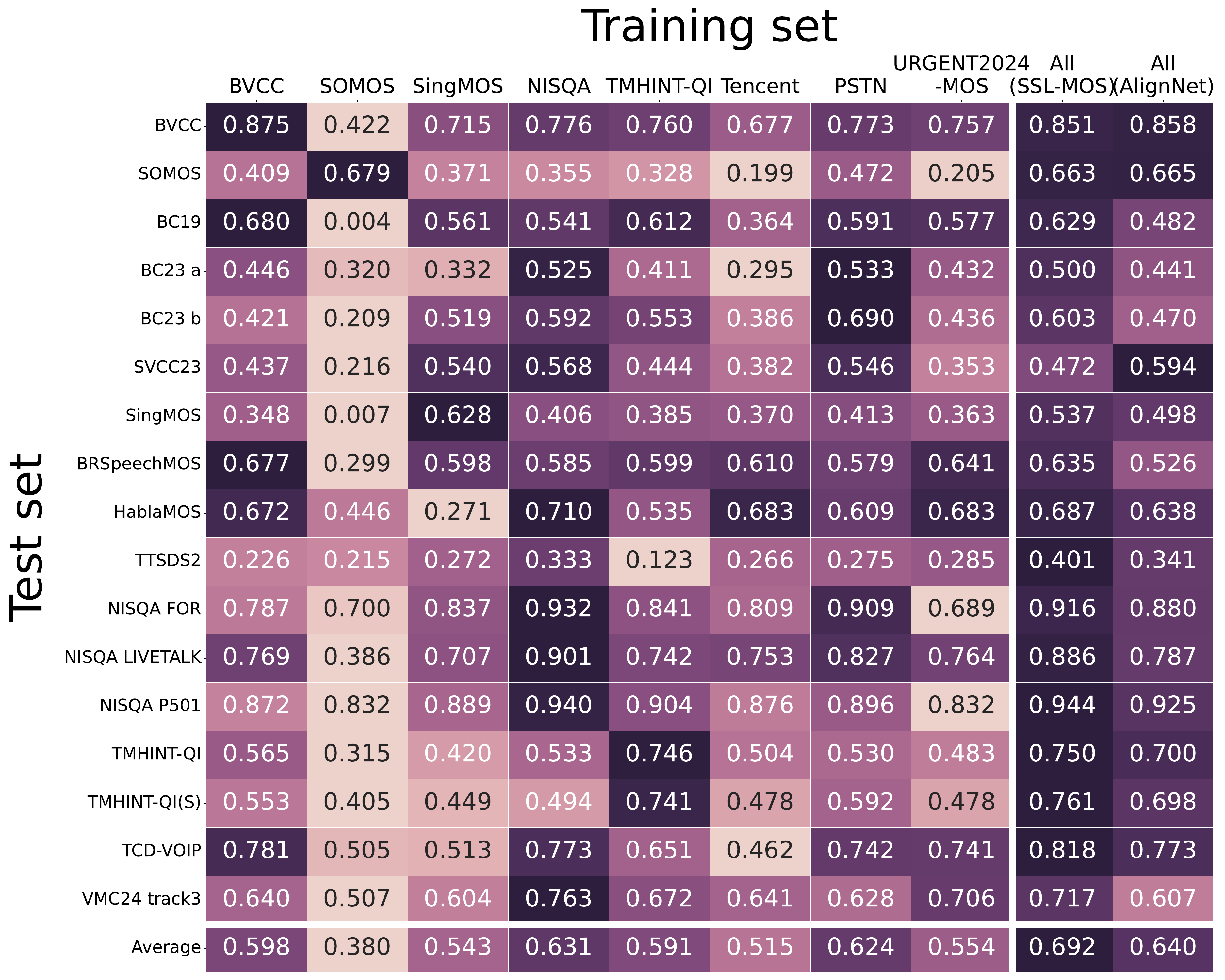}
    \caption{Utterance-level SRCC}
    \label{fig:single-utt-srcc}
  \end{subfigure}
  \hfill
  \begin{subfigure}{0.49\textwidth}
    \centering
    \includegraphics[width=\linewidth]{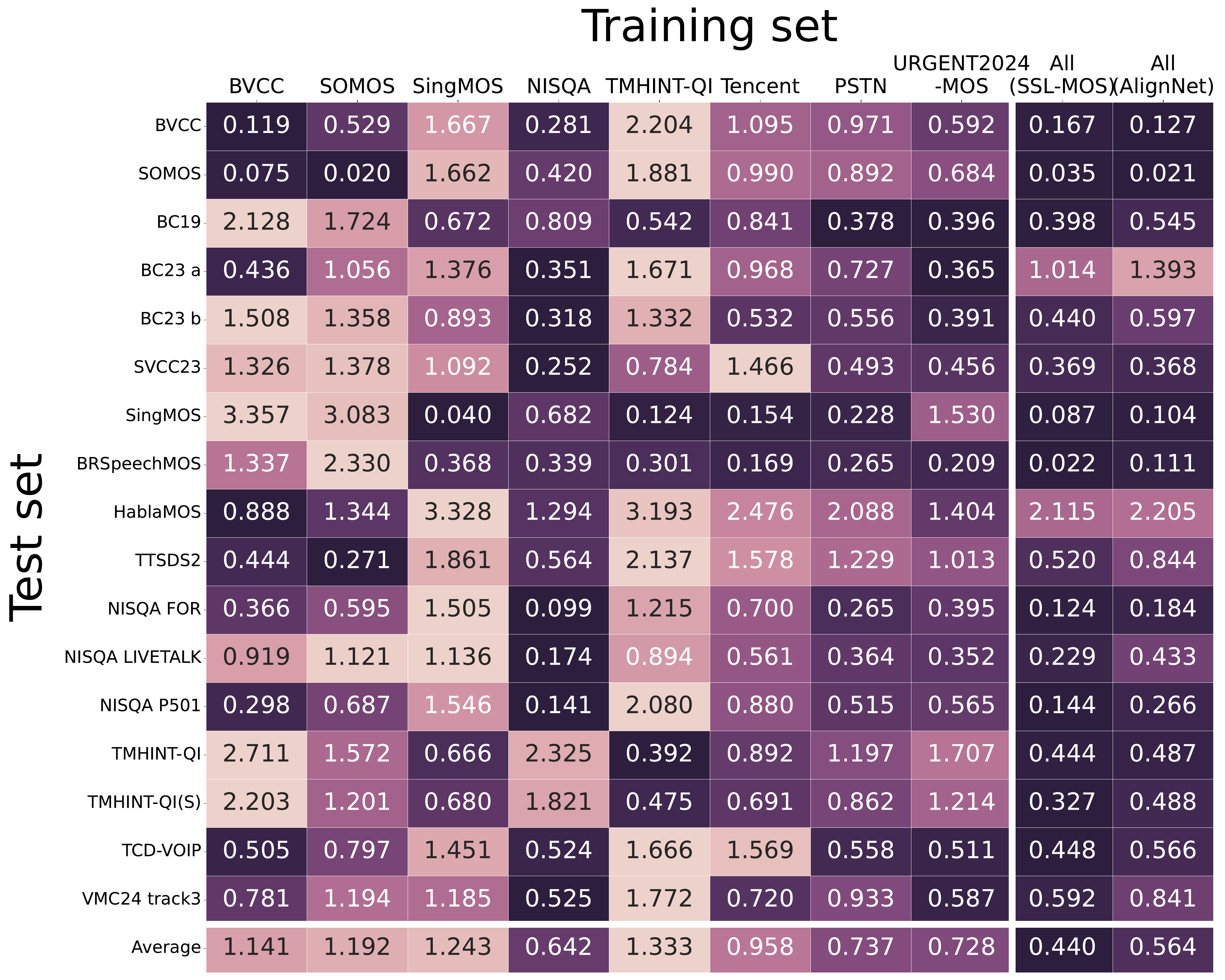}
    \caption{Utterance-level MSE}
    \label{fig:single-utt-mse}
  \end{subfigure}
  \caption{
  \rev{
  Heatmap results of the single dataset training and multiple dataset training experiments. For both figures, the higher the color intensity, the higher the performance. For SRCC, the higher the color intensity, the higher the performance. For MSE, the lower the color intensity, the better the performance. Color intensity is normalized row-wise to compare models on a specific test set.
  }
  }
  \label{fig:single-utt}
\end{figure*}

\section{Experimental settings}

\subsection{Training settings}

All experiments were conducted using the SHEET toolkit \cite{sheet}. For all experiments, following \cite{ssl-mos}, we set the training batch size to 16 and used the SGD optimizer with an initial learning rate of 0.001 and a momentum of 0.9.
Early stopping with patience was applied: at fixed intervals, we computed utterance-level SRCC (which will be explained in Sec.\ref{ssec:evaluation-metrics}) on the validation set and retained the five best checkpoints. Training stopped when none of these were updated after 20 validations or when the maximum number of training steps was reached. For single-dataset experiments described in Sec.\ref{sec:exp-single-dataset}, validation was performed every 100 steps with a cap of 100,000 steps. For multi-dataset experiments described in Sec.~\ref{sec:exp-multiple-dataset}, validation was conducted every 1,000 steps with a cap of 200,000 steps.

For single-dataset training experiments described in Sec.~\ref{sec:exp-single-dataset}, the SSL-MOS model was used, and for multi-dataset training experiments in Sec.~\ref{sec:exp-multiple-dataset}, we used SSL-MOS and AlignNet. For both models, we considered the choice of the SSL model as a hyperparameter and chose the best among the following three: HuBERT large (HuBERT-L) \cite{hubert}, wav2vec 2.0 large (wav2vec 2.0 L) \cite{wav2vec2}, and WavLM large (WavLM-L) \cite{wavlm}. We always used the last layer. Since all the above-mentioned SSL models took 16 kHz waveforms as input, all input speech samples were resampled to 16 kHz.

\subsection{Evaluation metrics}
\label{ssec:evaluation-metrics}

We report two metrics: the \textbf{utterance-level mean square error (UTT-MSE)} and the \textbf{utterance-level Spearman’s rank correlation coefficient (UTT-SRCC)}. For readers interested in the implementation, we provide a code snippet online\footnote{\url{https://gist.github.com/unilight/883726c94640cca1f4d4068e29c3d20f}}.

One might ask \textbf{why SRCC is adopted instead of the linear correlation coefficient} (LCC, also known as Pearson’s correlation coefficient, PCC), which is commonly used in the literature on SQA \cite{nisqa}. The reason is that SRCC captures ranking relationships more effectively, which we consider particularly important in SQA. For example, in challenges such as BC or VCC, organizers and participants are more likely to be interested in the relative ranking of systems.

Another question may concern \textbf{why metrics are calculated at the utterance level rather than at the system level}, as in the VMCs. In many cases, such as speech distorted with background noise, the noise type and level can vary across utterances, making it difficult and less meaningful to define a ``system''. Moreover, when system-level metrics are applicable (e.g., in the comparison of TTS systems), we usually observe a strong correlation between system-level and utterance-level results. For this reason, in this paper, we adopt utterance-level metrics consistently across all speech types.

\begin{figure*}[t]
  \centering

  \begin{subfigure}{0.24\textwidth}
    \centering
    \begin{subfigure}{\linewidth}
      \centering
      \includegraphics[width=\linewidth]{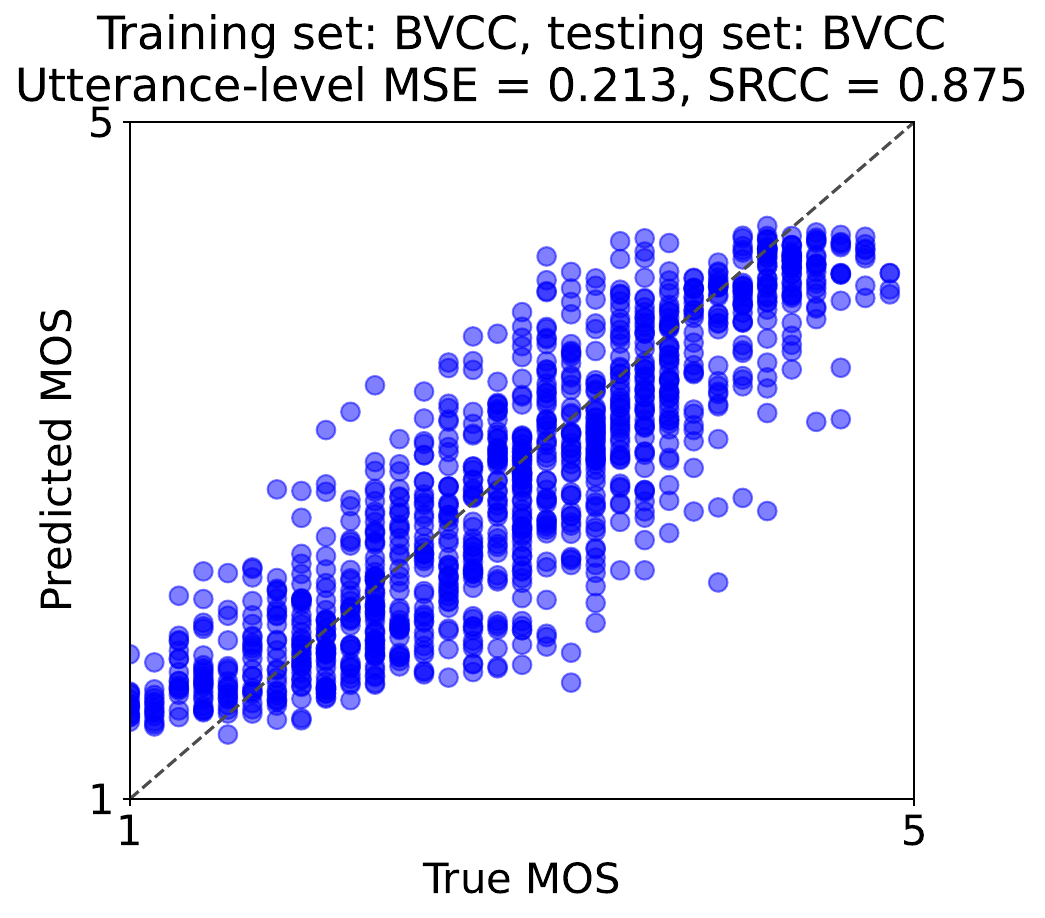}
    \end{subfigure}

    \vspace{2mm}

    \begin{subfigure}{\linewidth}
      \centering
      \includegraphics[width=\linewidth]{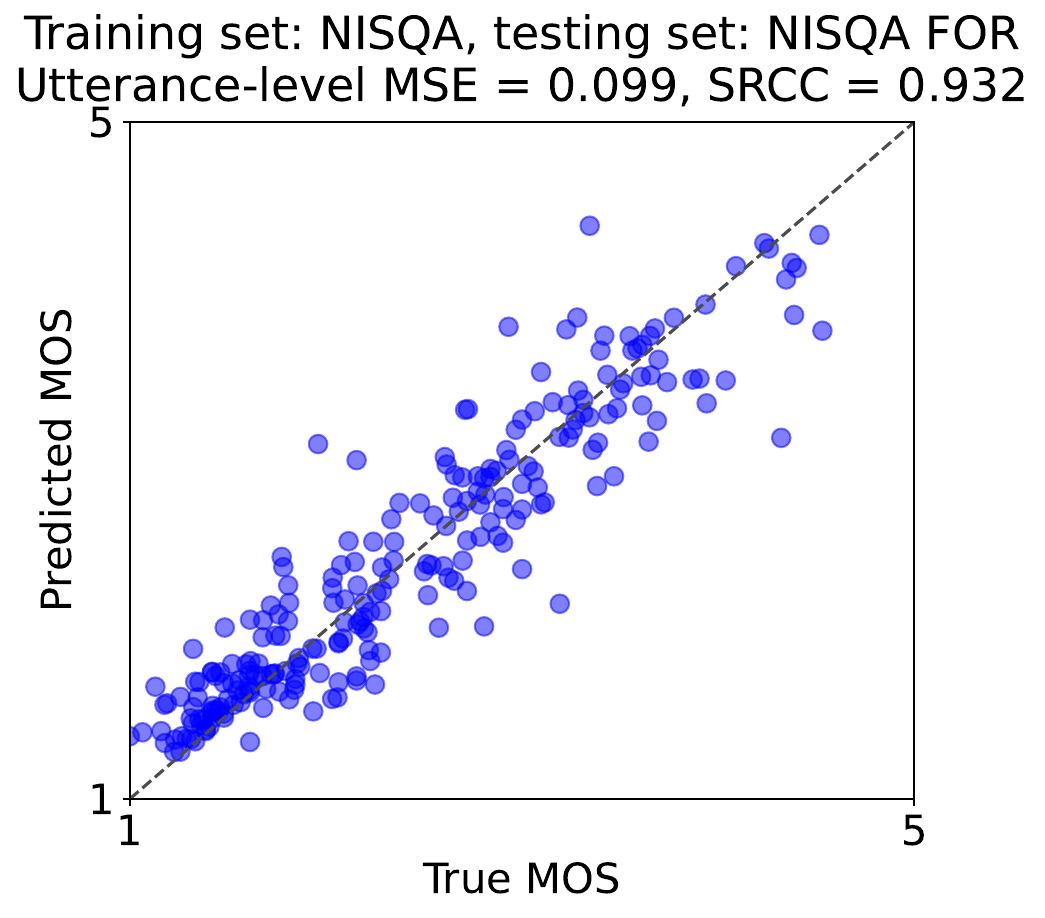}
    \end{subfigure}

    \caption{Low MSE, high SRCC}
    \label{fig:low-mse-high-srcc}
  \end{subfigure}%
    \begin{subfigure}{0.24\textwidth}
    \centering
    \begin{subfigure}{\linewidth}
      \centering
      \includegraphics[width=\linewidth]{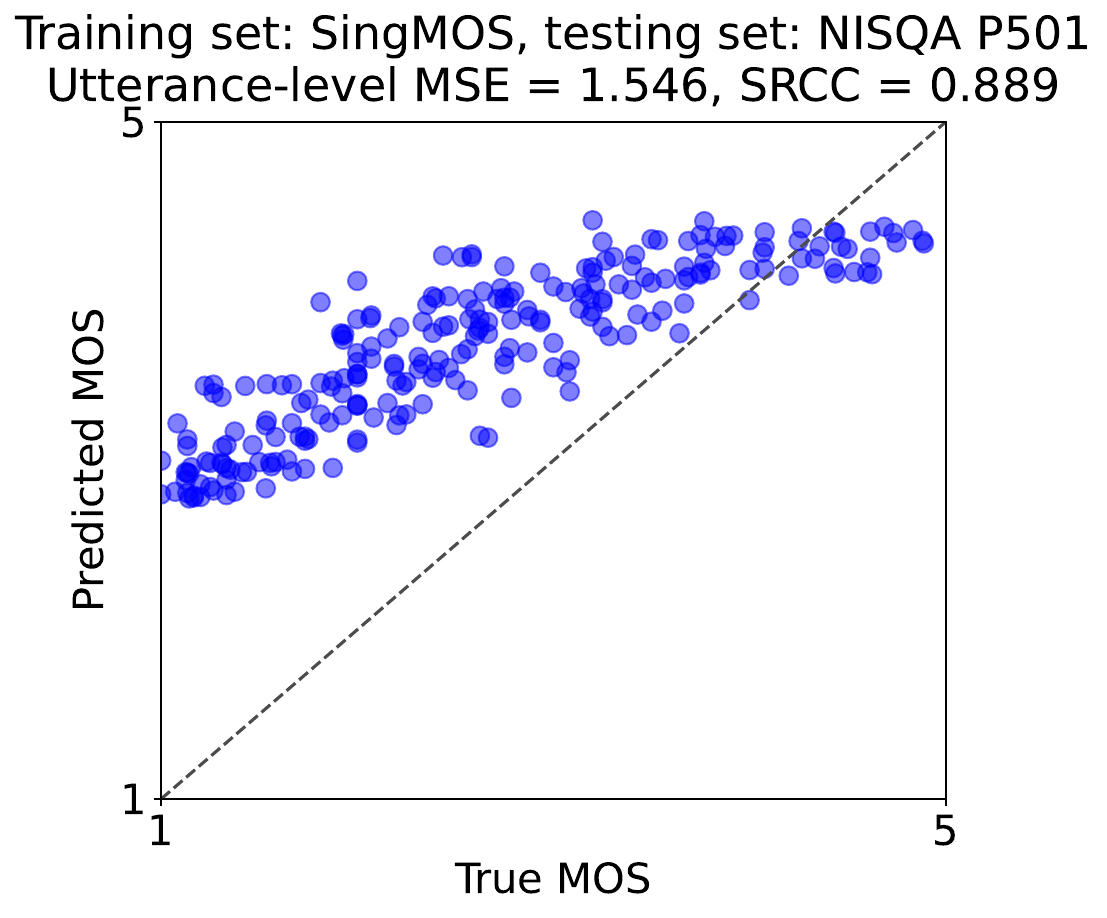}
    \end{subfigure}

    \vspace{2mm}

    \begin{subfigure}{\linewidth}
      \centering
      \includegraphics[width=\linewidth]{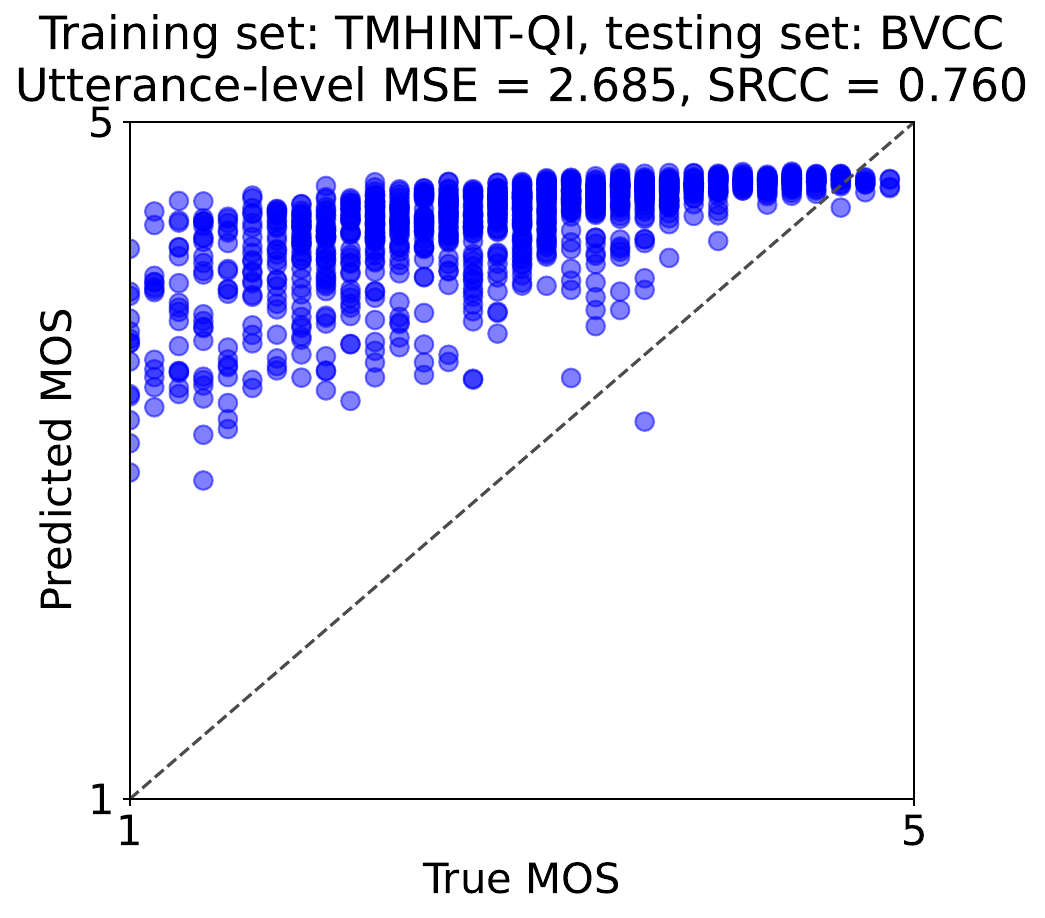}
    \end{subfigure}

    \caption{High MSE, high SRCC}
    \label{fig:high-mse-high-srcc}
  \end{subfigure}%
  \begin{subfigure}{0.24\textwidth}
    \centering
    \begin{subfigure}{\linewidth}
      \centering
      \includegraphics[width=\linewidth]{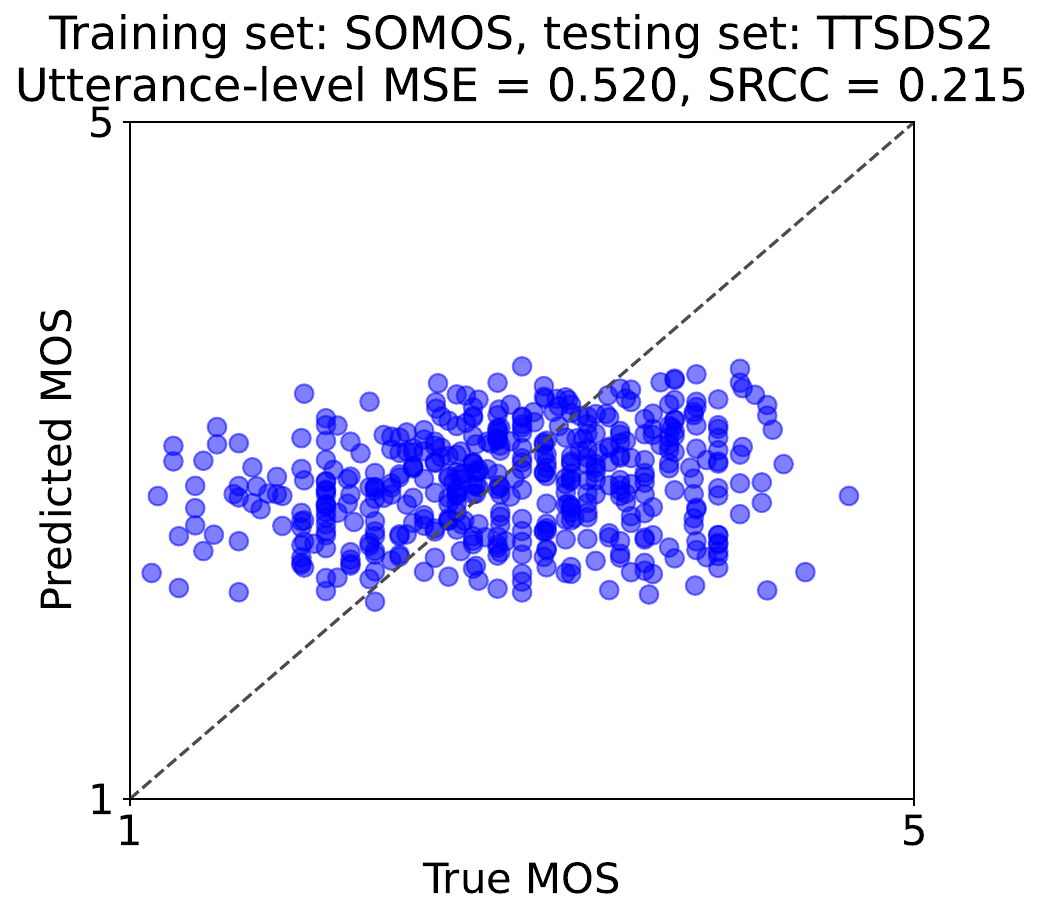}
    \end{subfigure}

    \vspace{2mm}

    \begin{subfigure}{\linewidth}
      \centering
      \includegraphics[width=\linewidth]{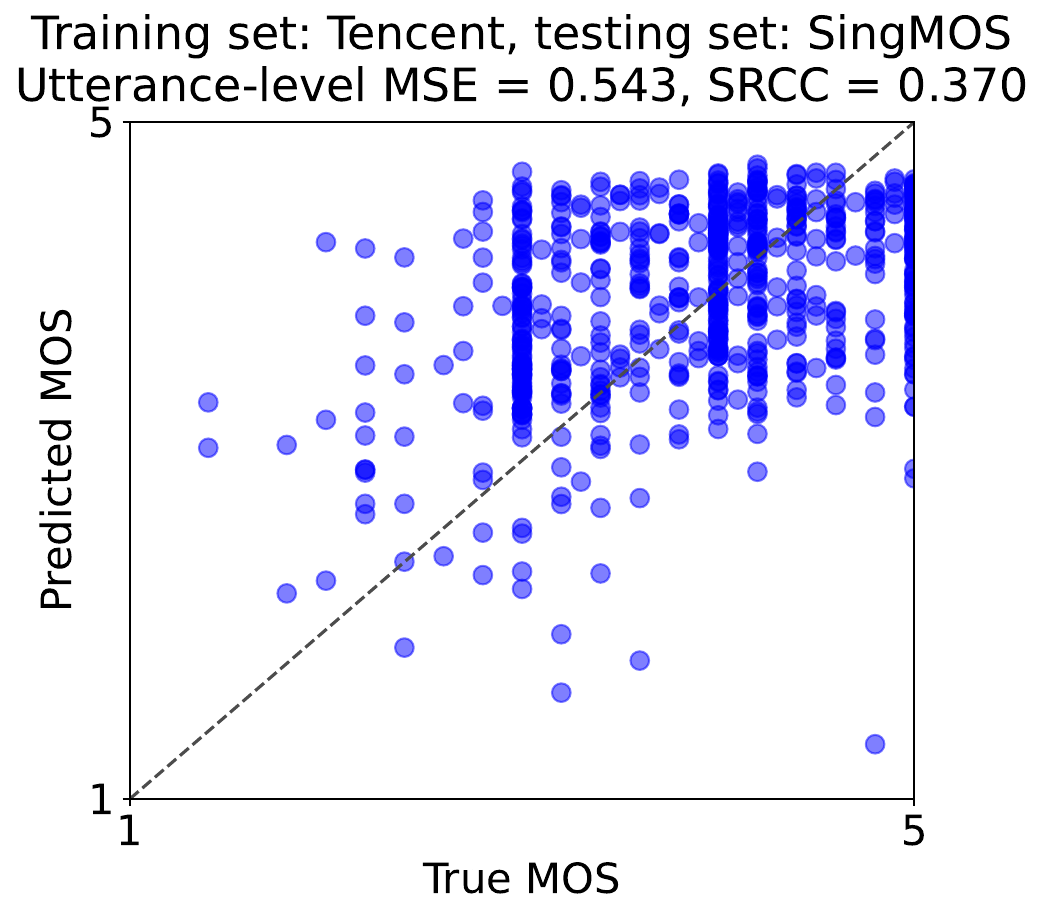}
    \end{subfigure}

    \caption{Low MSE, low SRCC}
    \label{fig:low-mse-low-srcc}
  \end{subfigure}%
  \begin{subfigure}{0.24\textwidth}
    \centering
    \begin{subfigure}{\linewidth}
      \centering
      \includegraphics[width=\linewidth]{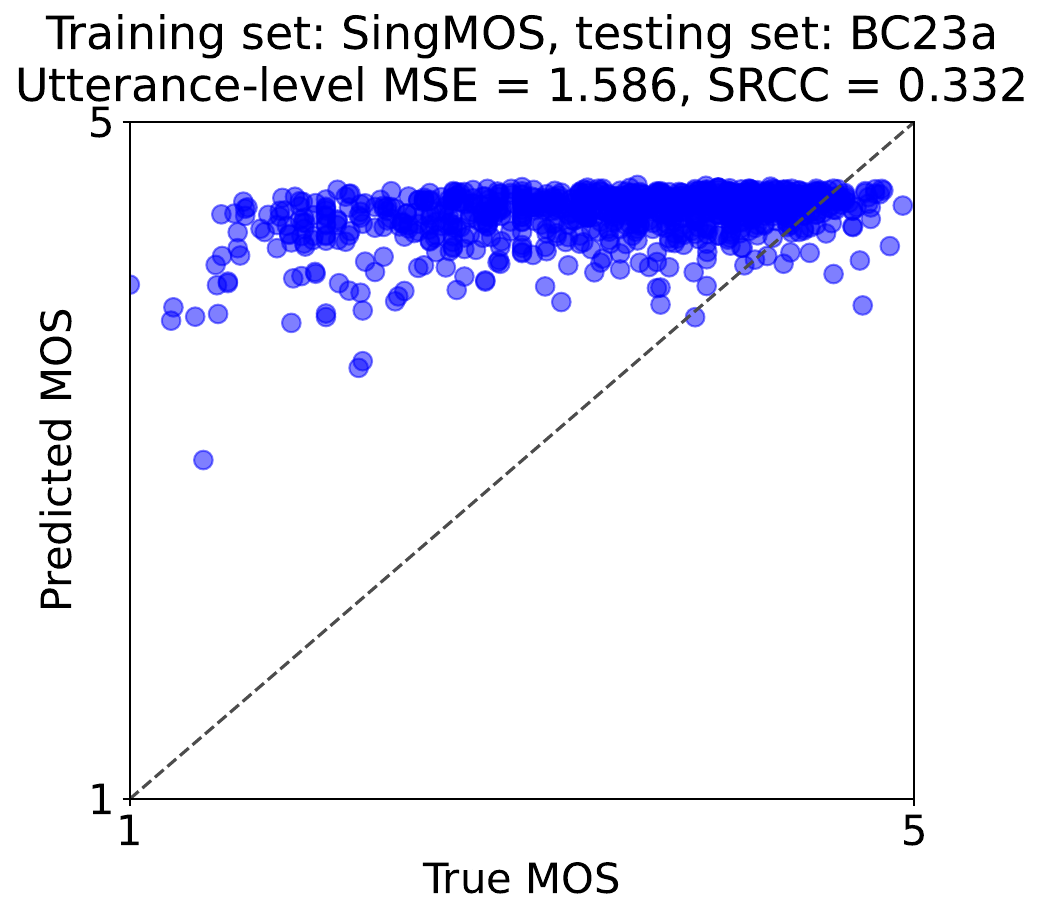}
    \end{subfigure}

    \vspace{2mm}

    \begin{subfigure}{\linewidth}
      \centering
      \includegraphics[width=\linewidth]{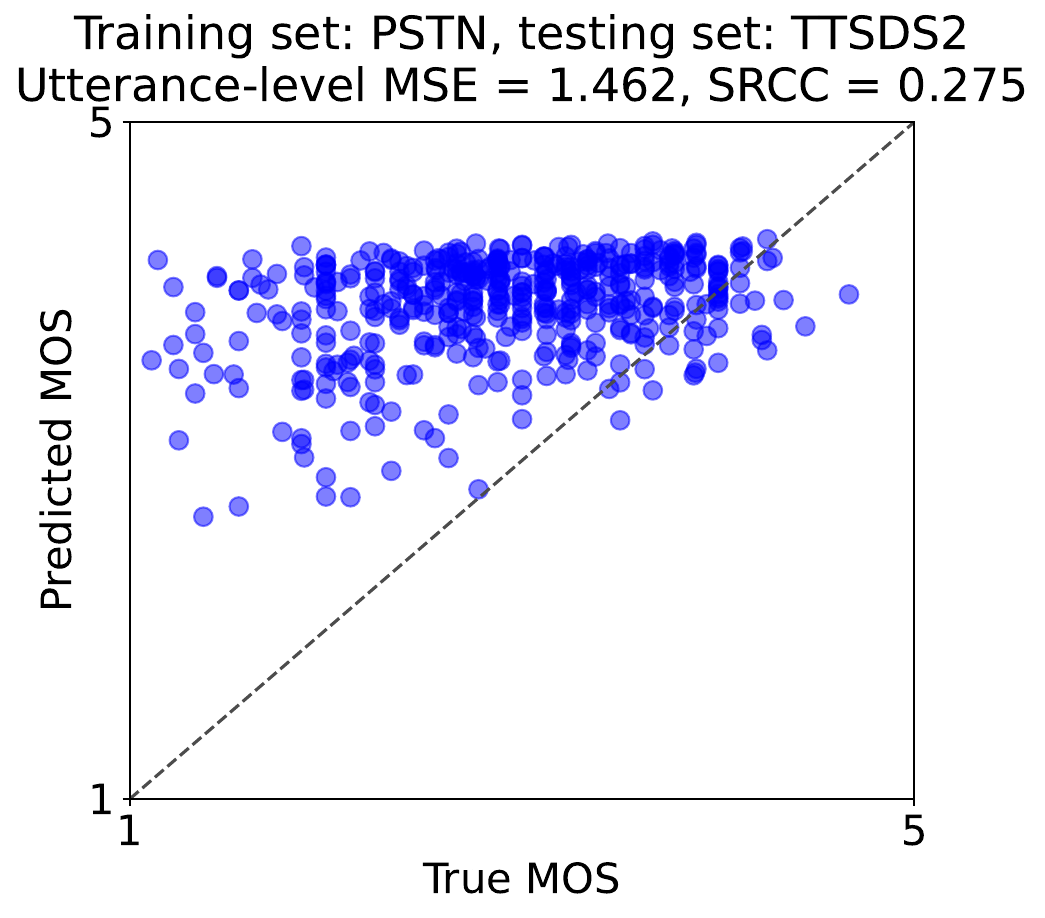}
    \end{subfigure}

    \caption{High MSE, low SRCC}
    \label{fig:high-mse-low-srcc}
  \end{subfigure}

  \caption{Sample scatter plots of true MOS against predicted MOS in different combinations of MSE and SRCC. All numbers are calculated at the utterance level.}
  \label{fig:utt-scatter}
\end{figure*}

\section{Single-dataset training experiments}
\label{sec:exp-single-dataset}

In this section, we trained SSL-MOS models using either one of the eight training sets in MOS-Bench listed in Table~\ref{tab:mos-bench-train}, and we benchmarked their performance on the 17 test sets in MOS-Bench listed in Table~\ref{tab:mos-bench-test}.

\subsection{Main results}
\label{ssec:exp-single-dataset-main-results}

Figure~\ref{fig:single-utt} shows the UTT-SRCC and UTT-MSE results of the single-dataset experiments in the form of heatmaps. For each test set, color intensity indicates the relative performance of a training set across all eight. For example, the first row of Fig.~\ref{fig:single-utt-srcc} shows UTT-SRCC on the BVCC test set. As expected, the model trained on BVCC achieved the highest performance, as whosn by the darkest cell.

From the UTT-SRCC results in Fig.~\ref{fig:single-utt-srcc}, SOMOS performed substantially worse than all other training sets. This outcome is unsurprising, as SOMOS mainly consists of single-speaker neural TTS samples with various prosodic inputs, whereas the vocoder and training dataset remained fixed. Such characteristics limited the quality variation, reducing the effectiveness of the model trained on this dataset. For the other datasets, no clear trend was observed, but NISQA consistently yielded the highest performance, followed by PSTN.

Regarding the UTT-MSE results in Fig.~\ref{fig:single-utt-mse}, BVCC, SOMOS, SingMOS, and TMHINT-QI performed poorly, each with an average UTT-MSE above 1.0 across test sets. TMHINT-QI was the weakest overall, with an average UTT-MSE of 1.333. Interestingly, SOMOS -- although the worst in UTT-SRCC -- was not the worst in UTT-MSE. The remaining four datasets performed well on most test sets, with NISQA again achieving the highest overall results.

The strong performance of NISQA carries an important implication: building an effective SSQA training set does not necessarily require collecting samples from a wide range of speech generation systems. Instead, \textbf{generating distorted speech with high diversity} may be more beneficial. In fact, we observed that synthetic training sets such as BVCC, SOMOS, and SingMOS did not perform notably higher on synthetic test sets, nor did distorted training sets outperform others on distorted test sets.

We are also interested in whether language plays a role, but found no clear trend. For example, Chinese training sets (SingMOS, TMHINT-QI, and Tencent) did not show superior performance on Chinese test sets (BC19, SingMOS, TMHINT-QI, and TMHINT-QI (S)). Finally, since NISQA and PSTN were the top two training sets in both UTT-SRCC and training size, we hypothesize that dataset size may be an important factor. We examine this hypothesis in Sec.~\ref{ssec:size-vs-performance}.

\subsection{Performance pattern analysis}

Given the contrasting trends when using UTT-SRCC and UTT-MSE as evaluation metrics as shown in Fig.~\ref{fig:single-utt}, we are interested in how predictions actually look like. For instance, if a model has not only a high UTT-SRCC but also a large UTT-MSE, what does it mean? To this end, we selected two representative cases for each scenario and plotted scatter plots of true MOS versus predicted MOS in Fig.~\ref{fig:utt-scatter}. A perfect system would show all points on the diagonal line, indicating that the predicted MOS values are identical to the true MOS. In the ideal case -- small MSE and large SRCC (Fig.~\ref{fig:low-mse-high-srcc}) -- most points lie close to the diagonal line, indicating that predictions are accurate for individual samples and rankings are preserved. This situation typically arises when training and test sets share similar distributions, often, though not always, in the in-domain setting.

Figure~\ref{fig:high-mse-high-srcc} illustrates the case of high SRCC but also large MSE. Here, points are far from the diagonal line, leading to large MSEs, yet they align along an oblique line, showing that the model preserves relative rankings, thus a high SRCC. In contrast, Fig.~\ref{fig:low-mse-low-srcc} shows the opposite: predictions cluster near the diagonal line, yielding a small MSE, but the points fail to form an oblique pattern. This indicates that the model predicts similar MOS values across samples, regardless of their true MOS, resulting in a low SRCC. Finally, Fig.~\ref{fig:high-mse-low-srcc} shows the worst case, where both metrics are poor: predictions deviate substantially from the diagonal line, and exhibit no clear ranking structure.

\begin{table}[t]
\scriptsize
\centering
\caption{
    \rev{
    Utterance-level SRCC and MSE results (mean $\pm$ standard deviation), which are calculated across all 17 test sets in MOS-Bench. ``All'' denotes training on the combination of all eight training sets in MOS-Bench. The best results for each metric are highlighted in bold.
    }
}
\label{tab:utt-srcc-mse-all}
\begin{tabular}{@{}ccc|cc@{}}
\toprule
Training set   & Model    & SSL model     & UTT-SRCC       & UTT-MSE        \\ \midrule
BVCC           & SSL-MOS  & WavLM-L       & 0.598 $\pm$ 0.192         & 1.141 $\pm$ 0.962         \\
SOMOS          & SSL-MOS  & HuBERT-L      & 0.380 $\pm$ 0.225          & 1.192 $\pm$ 0.743          \\
SingMOS        & SSL-MOS  & wav2vec 2.0 L & 0.543 $\pm$ 0.181          & 1.243 $\pm$ 0.735          \\
NISQA          & SSL-MOS  & WavLM-L       & 0.631 $\pm$ 0.193 & 0.642 $\pm$ 0.617 \\
TMHINT-QI      & SSL-MOS  & wav2vec 2.0 L & 0.591 $\pm$ 0.203          & 1.333 $\pm$ 0.845          \\
Tencent        & SSL-MOS  & wav2vec 2.0 L & 0.515 $\pm$ 0.201          & 0.958 $\pm$ 0.568          \\
PSTN           & SSL-MOS  & wav2vec 2.0 L & 0.624 $\pm$ 0.168          & 0.737 $\pm$ 0.472          \\
\begin{tabular}[c]{@{}c@{}}URGENT\\ 2024-MOS\end{tabular} & SSL-MOS  & WavLM-L       & 0.554 $\pm$ 0.189          & 0.728 $\pm$ 0.464          \\ \midrule
All            & SSL-MOS  & WavLM-L       & \textbf{0.692 $\pm$ 0.160} & \textbf{0.440 $\pm$ 0.497} \\
All            & AlignNet & HuBERT-L      & 0.640 $\pm$ 0.168          & 0.564 $\pm$ 0.544          \\ \bottomrule
\end{tabular}
\end{table}

\section{Multi-dataset training experiments}
\label{sec:exp-multiple-dataset}

In this section, we aim to investigate the effectiveness and outcomes of training SSQA models (specifically, SSL-MOS and AlignNet) using a combination of multiple training sets. To this end, we trained models using the combination of the eight training sets in MOS-Bench as listed in Table~\ref{tab:mos-bench-train}. Throughout this section, we denote the combined dataset as ``All''. Then, we benchmarked the models trained with ``All'' on the 17 test sets in MOS-Bench listed in Table~\ref{tab:mos-bench-test}.

\subsection{Main results}
\label{ssec:exp-multi-dataset-main-results}

\rev{
We are mainly interested in whether combining multiple datasets can bring \textbf{overall} improvements compared with single-dataset training, so instead of showing test-set-wise performance breakdown in heatmaps as in the previous section, we report UTT-SRCC and UTT-MSE averaged across all 17 test sets, along with their standard deviations, in Table~\ref{tab:utt-srcc-mse-all}. 
}

\rev{
To rigorously evaluate the effectiveness of multi-dataset training, we performed a paired t-test across the 17 test sets to compare the SSL-MOS model trained on the pooled ``All'' dataset against the best-performing single-dataset model (NISQA). The analysis reveals that the pooled model achieves a statistically significant improvement in UTT-SRCC, increasing from 0.631 to 0.692 with $p = 0.039$. Regarding UTT-MSE, although the average error numerically decreased from 0.642 to 0.440, the test yielded a p-value of 0.225. Although the absolute error reduction is numerically large, we suspect that its statistical significance is affected by the scale shifts between different MOS datasets. Importantly, however, the pooled training strategy notably reduces the standard deviation for both SRCC ($\pm$0.193 to $\pm$0.160) and MSE ($\pm$0.617 to $\pm$0.497). This narrowed variance demonstrates that multi-dataset training yields a more consistent and robust estimator across diverse acoustic domains.
}

On the other hand, the AlignNet model was outperformed by the SSL-MOS model. We will discuss this outcome in Sec.~\ref{ssec:why-does-not-alignnet-improve}.

\begin{figure}[t]
  \centering
  \begin{subfigure}{0.45\textwidth}
    \centering
    \includegraphics[width=\linewidth]{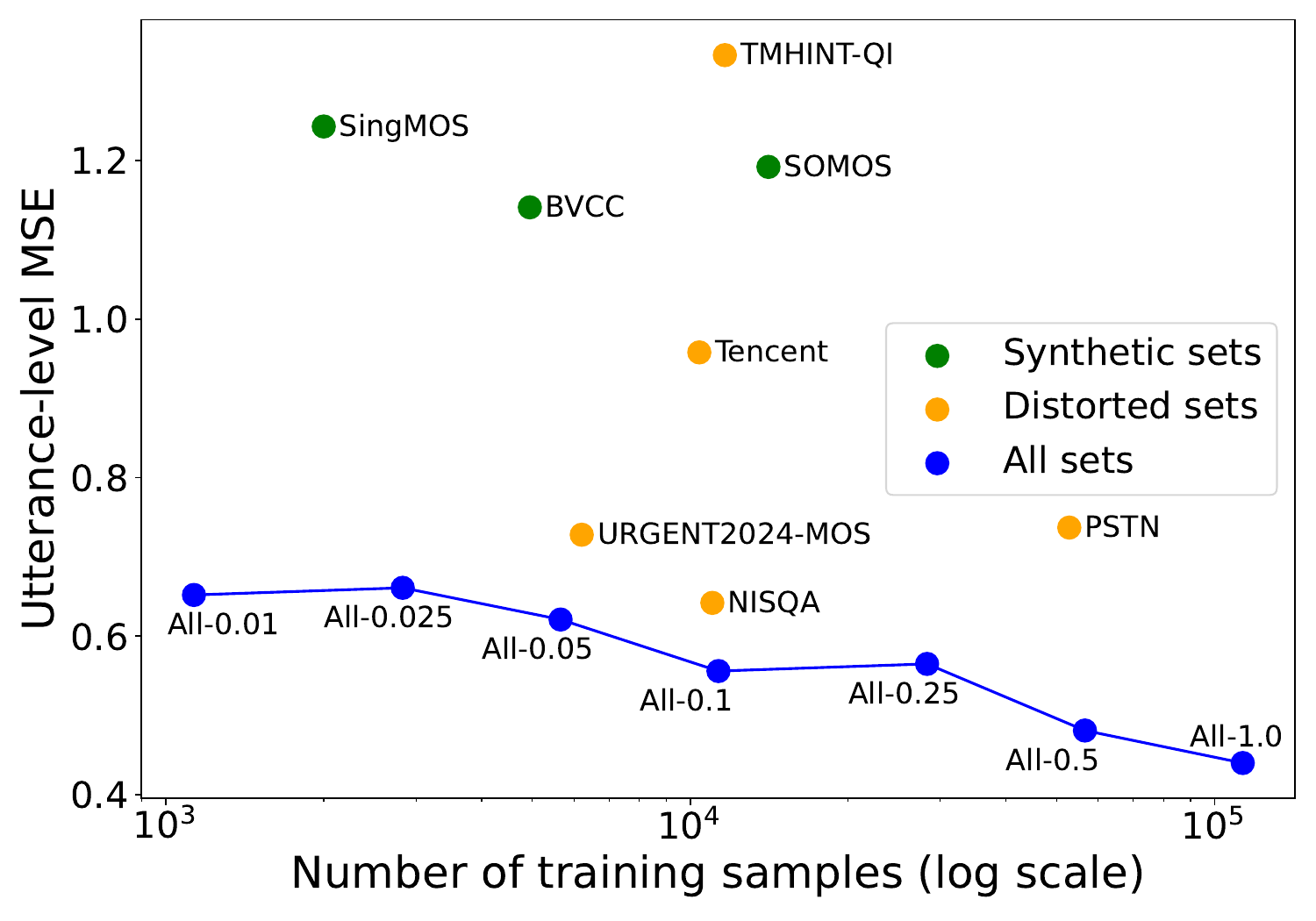}
    \caption{Utterance-level MSE}
    \label{fig:training-sample-vs-utt-mse}
  \end{subfigure}

  \vspace{10pt}
  
  \begin{subfigure}{0.45\textwidth}
    \centering
    \includegraphics[width=\linewidth]{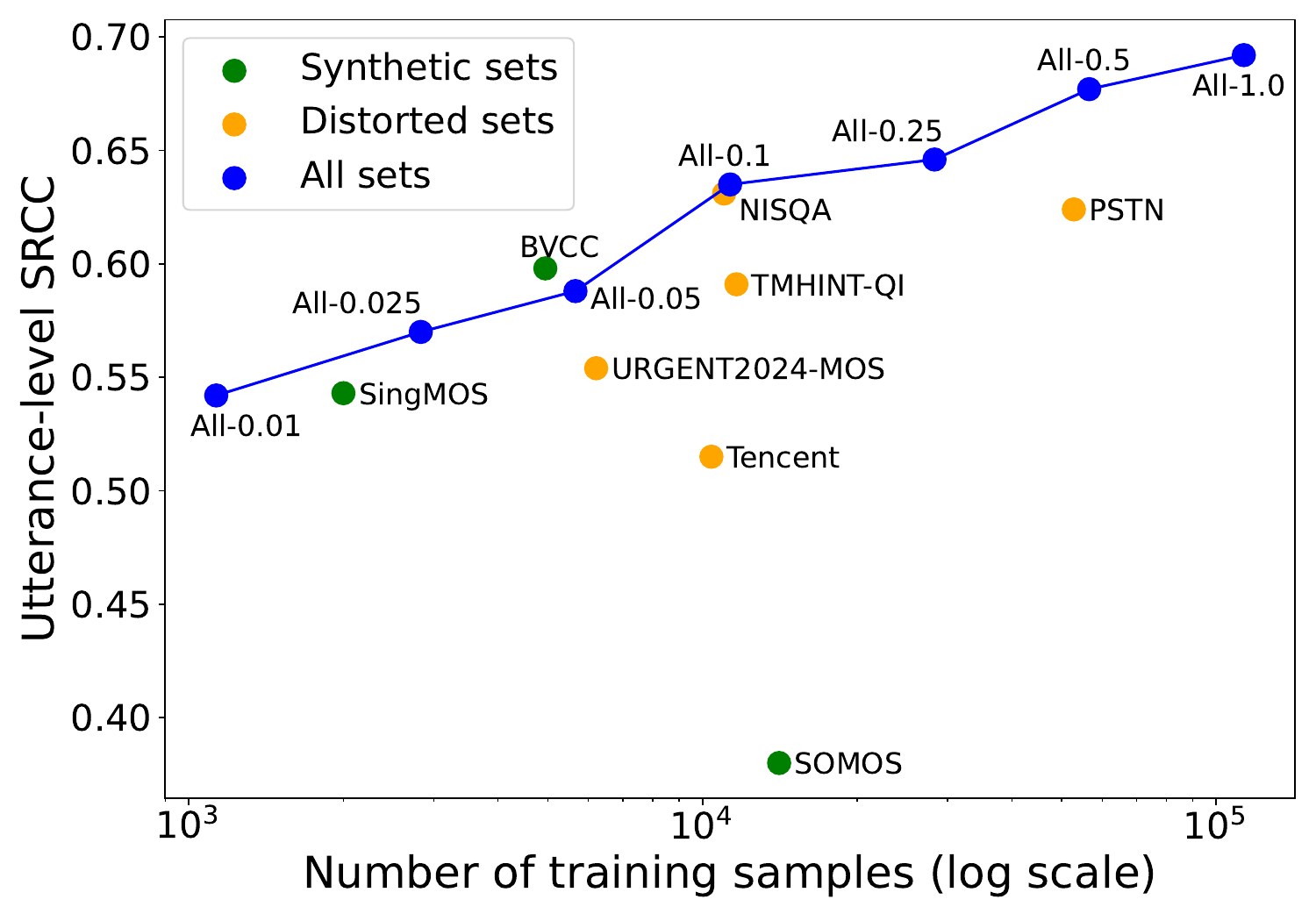}
    \caption{Utterance-level SRCC}
    \label{fig:training-sample-vs-utt-srcc}
  \end{subfigure}
  
  \caption{Scatter plot between training data size (log scale) and averaged utterance-level MSE/SRCC across all test sets. ``All-$X$'' indicates a subset of the full training data, where $X$ denotes the fraction used. For example, ``All-0.5'' corresponds to using half of the full training set.}
  \label{fig:training-sample-vs-utt}
\end{figure}

\subsection{Does increasing training data size always lead to higher performance?}
\label{ssec:size-vs-performance}

It is natural to attribute a model’s success primarily to the size of its training data. As shown in the single-dataset results in Sec.~\ref{ssec:exp-single-dataset-main-results}, the two highest-performing models were trained on NISQA and PSTN, both of which are relatively large training sets in MOS-Bench. Similarly, in Sec.~\ref{ssec:exp-multi-dataset-main-results}, the ``All'' dataset -- containing 113,194 samples, more than twice the size of the largest training set, PSTN -- \rev{achieved the most stable overall generalization.}. However, we hypothesize that \emph{diversity}, not just size, also plays a vital role in improving performance.

To test this hypothesis, we conducted an ablation study by gradually reducing the size of the ``All'' dataset. Specifically, we randomly sampled fractions of the data and trained SSL-MOS models on the resulting subsets. The sampling ratios were 0.5, 0.25, 0.1, 0.05, 0.025, and 0.01. The smallest subset contained only 1,131 samples -- fewer than those in SingMOS, the smallest dataset in MOS-Bench. Figure~\ref{fig:training-sample-vs-utt} shows the results as scatter plots, where the x-axis (log scale) represents the number of training samples and the y-axis shows utterance-level MSE and SRCC.

We first note that, in the single-dataset training setting, a larger dataset does not necessarily guarantee higher performance. For example, SOMOS, despite being the second-largest dataset, performed the worst in terms of UTT-SRCC. Similarly, PSTN, the largest dataset, underperformed compared with NISQA and URGENT2024-MOS in terms of UTT-MSE, and was also inferior to NISQA in terms of UTT-SRCC.

Turning to the ``All'' subsets, Fig.~\ref{fig:training-sample-vs-utt-mse} shows that in terms of UTT-MSE, reduced versions of ``All'' consistently outperformed training sets of comparable size. For instance, among training sets with around 10,000 samples, ``All-0.1'' achieved the lowest UTT-MSE, outperforming TMHINT-QI, SOMOS, Tencent, and NISQA. A similar trend is evident in terms of UTT-SRCC in Fig.~\ref{fig:training-sample-vs-utt-srcc}, with one exception, that is, BVCC achieved a slightly higher UTT-SRCC than ``All-0.05,'' despite having fewer training samples.

\rev{
These results demonstrate that \textbf{training data size alone does not guarantee robust generalization}. Larger datasets can still yield high variance or subpar ranking ability if they lack sufficient variability. In contrast, combining multiple datasets -- even when downsampled -- consistently yields improved rank correlation and reduced performance variance, emphasizing that \emph{diversity} in training data is just as critical as size for stable OOD prediction.
}

\begin{figure*}[t]
    \centering
    \includegraphics[width=0.8\textwidth]{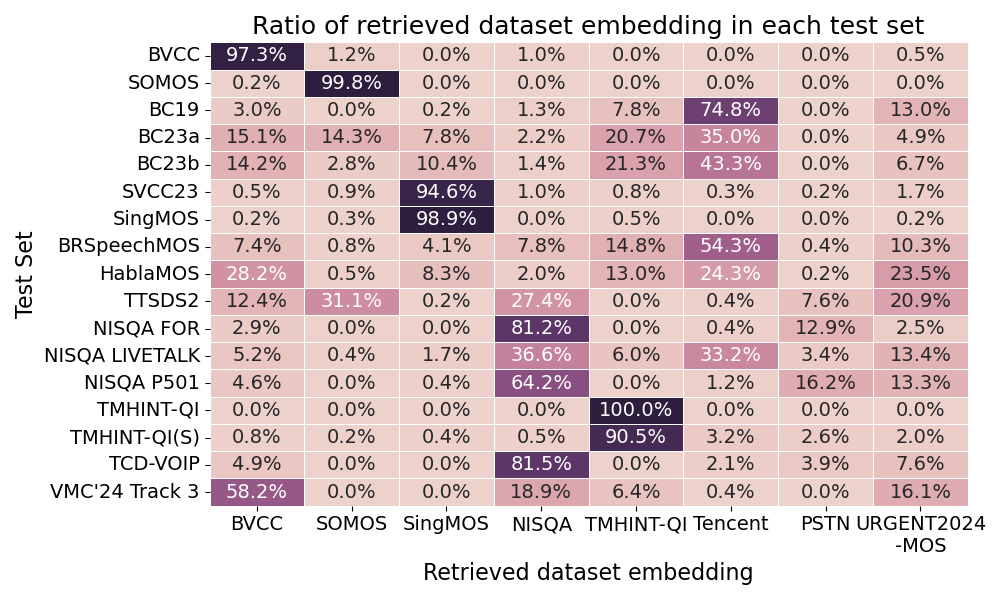}
    \caption{Ratio of the retrieved dataset embedding during inference with each test set using AlignNet trained with the eight training sets in MOS-Bench.}
    \label{fig:retrieval-id}
\end{figure*}

\begin{figure}[t]
  \centering
  \begin{subfigure}{0.4\textwidth}
    \centering
    \includegraphics[width=\linewidth]{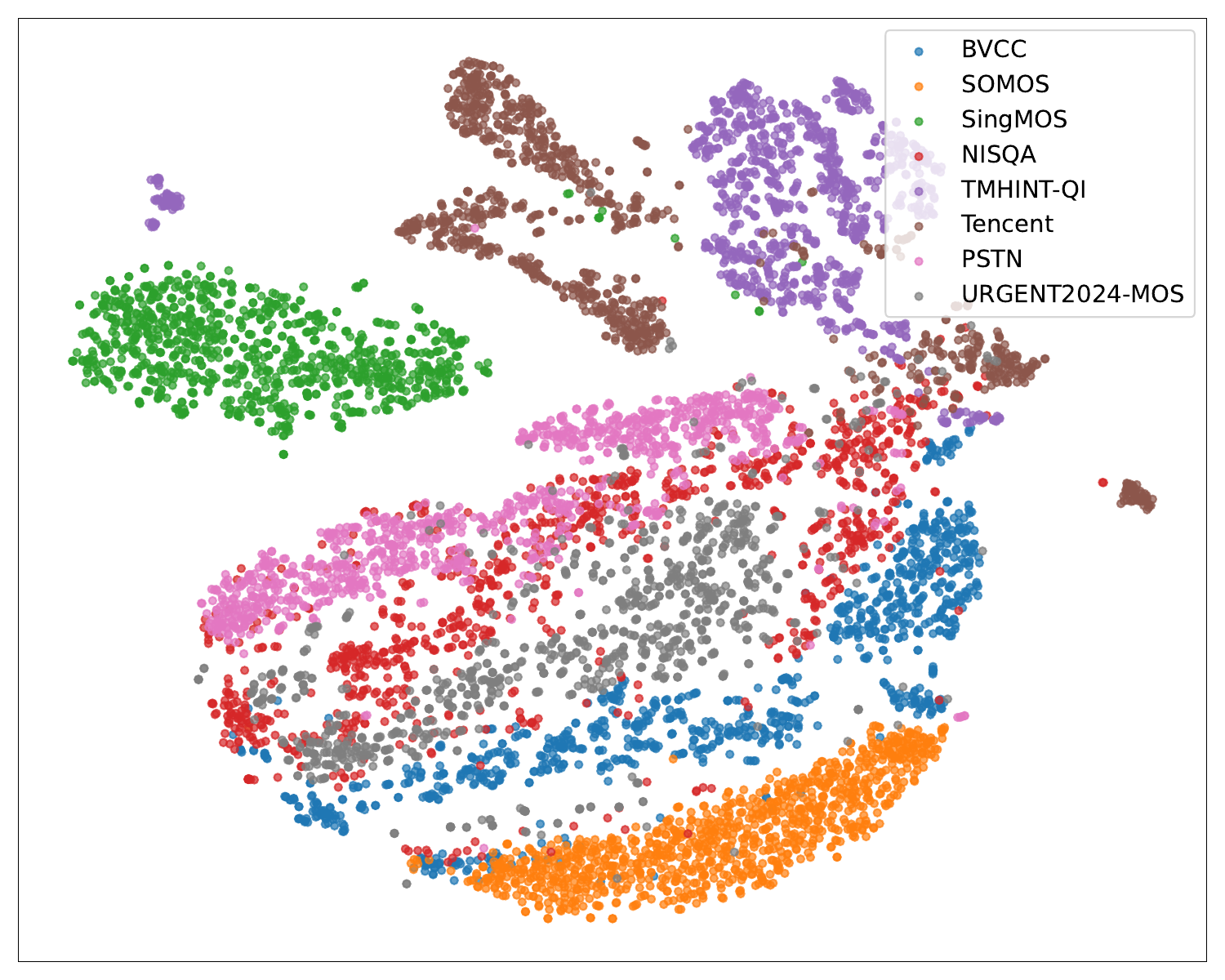}
    \caption{SSL-MOS}
    \label{fig:tsne-ssl-mos}
  \end{subfigure}
  \begin{subfigure}{0.4\textwidth}
    \centering
    \includegraphics[width=\linewidth]{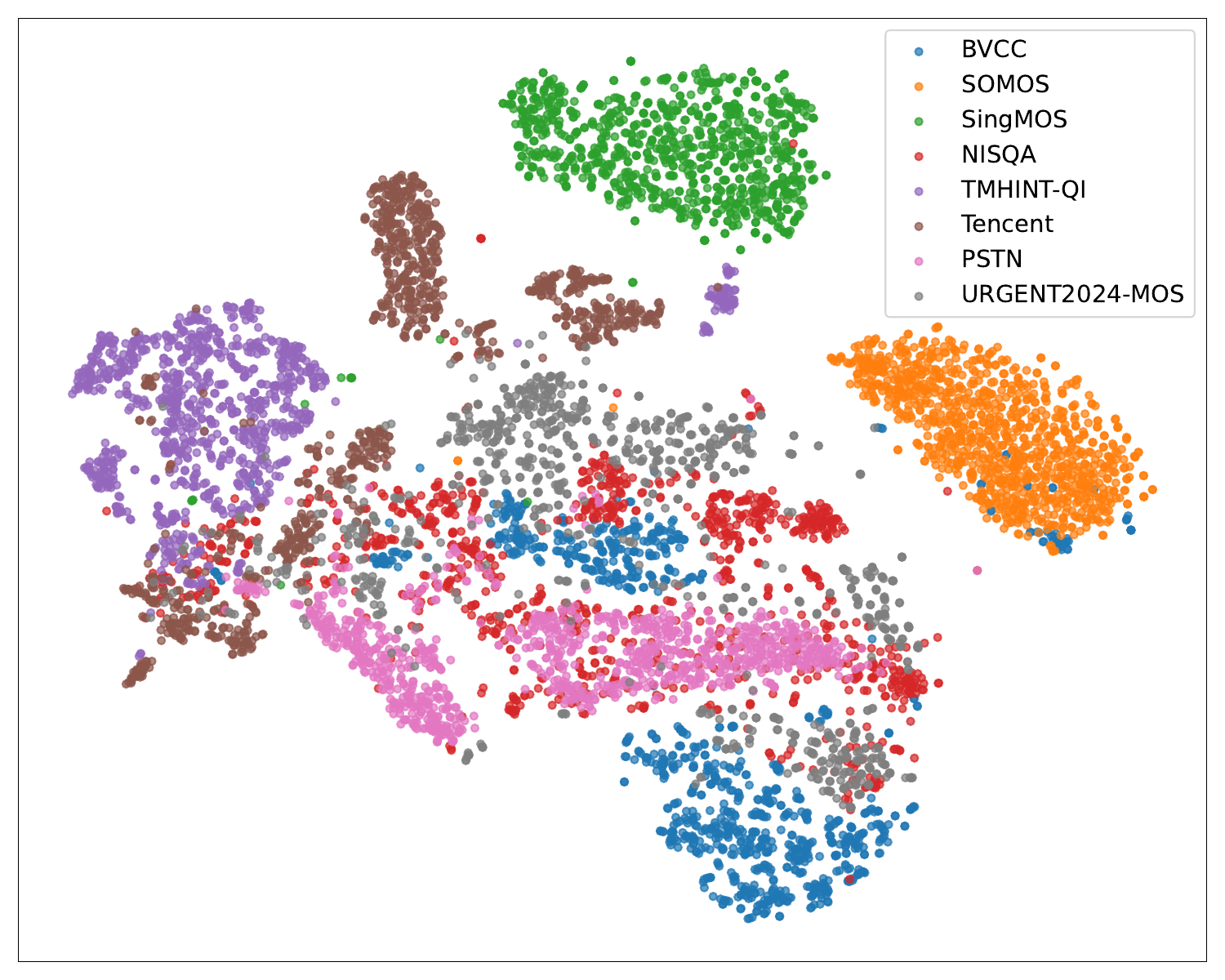}
    \caption{AlignNet}
    \label{fig:tsne-alignnet}
  \end{subfigure}
  
  \caption{t-SNE visualization of SSL representations extracted from 100 randomly sampled utterances per test set.}
  \label{fig:tsne}
\end{figure}

\subsection{Why does AlignNet not improve?}
\label{ssec:why-does-not-alignnet-improve}

We now analyze why AlignNet did not yield improvements over SSL-MOS, as shown in Table~\ref{tab:utt-srcc-mse-all}. A straightforward explanation is that the problem AlignNet was designed to address -- namely, the corpus effect -- may not be prominent in the current MOS-Bench training set collection. That being said, the corpus effect is inherently difficult to prove for existence directly. Recall that the corpus effect refers to the same type of speech receiving different quality scores across listening tests. To confirm its presence, a substantial amount of overlapping speech across multiple listening tests would be required, which is not available in the current eight training sets. Thus, our results can only \textit{indirectly} suggest that the corpus effect might not be a major issue. On this basis, we consider another approach, that is, \textbf{analyzing the behavior of AlignNet itself}.

\subsubsection{Analysis of the retrieved dataset embeddings}

We first revisit the \textit{dataset embedding retrieval} mechanism, in which, during inference, the dataset embedding of the nearest neighbor to the input sample is retrieved. Ideally, \textbf{the retrieved embedding should come from a dataset whose domain is close to the input sample}. For instance, if the input is a singing voice, the embedding of SingMOS should be selected. To examine this, we computed the distribution of retrieved embeddings across test sets, as shown in Fig.~\ref{fig:retrieval-id}.

The results reveal both expected and unexpected patterns. For test sets such as BVCC, SOMOS, SingMOS, NISQA, and TMHINT-QI, the embedding from the corresponding training set was retrieved most frequently, which aligns with the in-domain scenario. However, the expected cross-domain behavior was not consistently observed: for example, in the BC19 test set which contains Chinese TTS samples, we expected that embeddings from BVCC or SOMOS (both are synthetic datasets) to be retrieved. Instead, the Tencent embedding dominated, likely because it also contains Chinese speech. In BC23 a and BC23 b, which consist of French TTS samples, embeddings from Tencent and TMHINT-QI were retrieved rather than BVCC or SOMOS. Similarly, for the VMC’24 Track 3 test set, which primarily includes noisy and enhanced speech, BVCC was retrieved in more than half of the cases.

\subsubsection{Visualizing the learned SSL representations}

Another expected behavior of AlignNet is that conditioning the decoder with dataset embeddings should promote dataset-invariant SSL representations. To evaluate this, we compared the SSL representation spaces of SSL-MOS and AlignNet, both trained on ``All’’. From each training set, we extracted SSL representations of 1,000 randomly sampled utterances, applied t-SNE for dimensionality reduction, and visualized the resulting embeddings in two dimensions.

The results are shown in Fig.~\ref{fig:tsne}. For SSL-MOS (Fig.~\ref{fig:tsne-ssl-mos}), clear separation between datasets is evident, although some meaningful relationships appear. For example, TMHINT-QI (purple) and Tencent (brown) -- both Chinese datasets -- cluster together. PSTN (pink), NISQA (red), and URGENT2024-MOS (gray) -- all English distorted datasets -- form a nearby group, whereas BVCC (blue) and SOMOS (orange) -- English synthetic datasets -- also cluster together. In contrast, the AlignNet space (Fig.~\ref{fig:tsne-alignnet}) shows more overlap: distorted datasets such as TMHINT-QI, Tencent, PSTN, NISQA, and URGENT2024-MOS are mixed together in the middle-left region, with some BVCC samples interspersed. Nonetheless, distinct dataset boundaries remain visible, suggesting that although dataset conditioning promotes greater invariance, it does not fully eliminate dataset-specific clustering.

Taken together, these analyses suggest that AlignNet only partially achieves the expected behaviors. Further refinement of SSL representation learning and a deeper understanding of how dataset embeddings interact with the latent space remain important directions for future work.

\section{Conclusion and Future Directions}

We presented MOS-Bench, a collection of SSQA datasets designed to benchmark the generalization ability of SSQA models. The currently released version consists of eight training sets and 17 test sets, covering a wide variety of speech types, languages, and sampling frequencies. We provided brief descriptions of each dataset and critically analyzed label quality. Our experiments demonstrated that models trained on certain commonly used datasets (including BVCC and NISQA) fail to generalize across the full range of test sets in MOS-Bench. Beyond this, three key findings emerged: (1) datasets with diverse distortions tend to be more effective than those primarily covering many speech synthesis systems; (2) multi-dataset training consistently improves generalization; and (3) variation in the training data, not just dataset size, is crucial for robust performance.

\rev{
MOS-Bench is an ongoing effort. As the SSQA research community continues to grow, we anticipate more initiatives focused on collecting and releasing new datasets, and we plan to update MOS-Bench accordingly to ensure it remains a relevant and valuable benchmark. Future directions for SSQA research include the following:
\begin{itemize}
    \item \textbf{Including datasets of other listening test protocols.} As mentioned in Sec.~\ref{ssec:dataset-quality-assessment}, MOS tests have been criticized in recent years, leading to increasing interest in other listening test protocols including preference-based tests \cite{hu23d_interspeech, speechjudge} and multidimensional evaluation \cite{aes}. Including such datasets may increase the reliability and explainability of SSQA models.
    \item \textbf{Data selection.} Despite the effectiveness of multiple dataset training, naively training with all possible data may be sub-optimal in general. However,  the goodness of the data at the sample or the dataset level should be defined in order to perform filtering. In other speech processing tasks such as speech recognition, ``good data'' is usually ``unseen data'' and can be determined on the basis of model confidence \cite{active-learning-asr}. In the context of SQA, since the label itself is noisy, goodness may be ill-defined. Nonetheless, this is still a promising direction worth exploring.
\end{itemize}
}

\section*{Acknowledgment}

This work was partly supported by JSPS KAKENHI Grant Number 25K00143 and JST AIP Acceleration Research JPMJCR25U5, Japan.
The authors would like to thank Jaden Pieper and Stephen Voran, the authors of the AlignNet paper, for providing references to various datasets used in their work, as well as for fruitful discussions. The authors would also like to thank Alejandro Sosa Welford, the provider of the HablaMOS dataset, for clarification of the dataset. The authors are grateful to all the authors and curators of every dataset in MOS-Bench for kindly open-sourcing them.


\bibliographystyle{IEEEtran}
\bibliography{IEEEabrv, ref}

\end{document}